\documentclass[lettersize,journal]{IEEEtran}
\usepackage{amsmath,amsfonts}
\usepackage{algorithmic}
\usepackage{algorithm}
\usepackage{array}
\usepackage[caption=false,font=normalsize,labelfont=sf,textfont=sf]{subfig}
\usepackage{textcomp}
\usepackage{stfloats}
\usepackage{url}
\usepackage{verbatim}
\usepackage{graphicx}
\usepackage[subtle]{savetrees} 
\usepackage{xcolor}
\usepackage{booktabs}
\usepackage[]{hyperref} 
\usepackage{multicol}
\usepackage{comment}
\usepackage{mathtools}
\usepackage{tcolorbox}
\usepackage{mdframed}
\usepackage[normalem]{ulem}
\usepackage[style=ieee,citestyle=numeric-comp,doi=false,isbn=false,url=false,eprint=false,maxnames=2,minnames=1]{biblatex}
\usepackage[english]{babel}
\usepackage{amsthm}
\usepackage{bm}

\newtheorem{prop}{Proposition}
\newtheorem{remark}{Remark}

\newtheorem{innerdefinition}{Definition}
\newenvironment{definition}[1][]
  {\begin{mdframed}[leftmargin=0pt,rightmargin=0pt,innerleftmargin=6pt,innerrightmargin=6pt,innertopmargin=6pt,innerbottommargin=6pt]\begin{innerdefinition}[#1]\normalfont}
  {\end{innerdefinition}\end{mdframed}}

\DeclarePairedDelimiter\ceil{\lceil}{\rceil}

\newcommand{\Pagg}{P_\text{agg}}

\newcommand{\C}{^\circ \text{C}}

\begin{document}

\title{
Power-Duration Characterization of Aggregated Thermostatically Controlled Loads via Reach and Hold Sets
%Characterizing the Demand Response Capability of Thermostatically Controlled Loads with Reach and Hold Sets
}
% Reaching the limits of 
\author{Mazen Elsaadany, Hamid R. Ossareh, and Mads R. Almassalkhi

        % <-this % stops a space
    \thanks{The authors are with the Department of Electrical and Biomedical Engineering at the University of Vermont, Burlington, VT. {\tt\{melsaada,hossareh,malmassa\}@uvm.edu}}%

\thanks{We acknowledge support from NSF Awards ECCS-2047306 and CMMI-2238424.}% <-this % stops a space
}

% The paper headers
\markboth{Journal of \LaTeX\ Class Files,~Vol.~14, No.~8, August~2021}%
{Shell \MakeLowercase{\textit{et al.}}: A Sample Article Using IEEEtran.cls for IEEE Journals}

\maketitle

\begin{abstract}
Aggregations of thermostatically controlled loads (TCLs), such as air conditioners, offer valuable flexibility to the power grid. The aggregate power consumption of a TCL fleet can be controlled by adjusting thermostat setpoints. An \textit{ex-ante} quantification of the flexibility that results from such setpoint change can inform grid operator decisions. This paper develops a rigorous, yet practical method to quantify flexibility in terms of the `reach-and-hold' set of TCL aggregations, which defines how much power can be shifted (reach) and for how long (hold). To quantify the reach-and-hold set, we employ a Markov-chain-based model of the TCL aggregation that captures second-order TCL dynamics, enabling accurate characterization of reach-and-hold sets. A tractable optimization problem is then formulated to numerically compute an inner approximation of these sets. Simulation results validate that our method accurately characterizes the fleet's flexibility and effectively controls its power consumption. Furthermore, a robustness analysis is carried out to investigate the effects of uncertainty in initial conditions and TCL parameters.
\end{abstract}


\section{Introduction}\label{sec:intro}
The increasing penetration of variable renewable energy (VRE) and rapidly growing load demand necessitates enhanced flexibility within the power grid. This flexibility must be provided by supply-side assets, such as conventional bulk generation and battery storage, or by distributed energy resources (DERs) through demand-side coordination~\cite{27_Mads_DR_definition}. Characterizing operational capability of these distributed, demand-side resources is the primary focus of this paper.
%Distributed energy resources (DERs) are a promising source of such flexibility, particularly when coordinated to modify their aggregate power consumption in response to system needs. 
Among demand-side DERs, thermostatically controlled loads (TCLs), such as air conditioners, heat pumps, and water heaters, are especially attractive because of their thermal inertia and large share of aggregate demand~\cite{lbnl_DRfut_2017}. When aggregated as a virtual power plant (VPP), TCLs can provide valuable grid services including demand response (DR) and frequency regulation. A key question, however, is how to characterize in advance the DR capability of a TCL fleet under a given control mechanism. Such an \emph{ex-ante} characterization is valuable because it provides grid operators and aggregators with a practical description of the fleet’s flexibility. Characterizing the complex aggregate dynamics with simple power-duration curves can help grid operators overcome reliability  concerns associated with the integration and operation of DR programs. 
% Related power-duration representations have been explored in ~\cite{power_duration_data_driven,power_duration_experimental,power_duration_novel_ON_OFF}.

Controlling the aggregate demand of TCLs generally falls into two distinct paradigms: 1) ON/OFF control, where devices are \emph{directly} switched ON/OFF, and 2) thermostat setpoint control, which \emph{indirectly} switches devices ON/OFF by adjusting their thermostat setpoints and devices switch ON/OFF based on the local thermostat logic and temperature state. Direct load control can provide a faster and more explicit aggregate response, but it generally requires more intrusive device-level actuation and explicit management of cycling, compressor lockout, and comfort constraints. In contrast, thermostat setpoint control is often easier to implement in practice and is consistent with how many residential DR programs operate today. This distinction is important because it determines the available flexibility of a TCL fleet and the appropriate methods used to characterize that flexibility.
% the appropriate flexibility model.
In the direct ON/OFF control setting, several works characterize aggregate flexibility using abstractions such as virtual batteries or approximations of the Minkowski sum of device-level flexibility sets~\cite{4_HeHao_2014aggregate,3_coffman_cycling,12_wang2020_flexibility_lock_time,13_wang2021_enhanced_VB,1_Geometric_minkowski_approx}. Previous power-duration representations of flexibility have been developed for TCLs under direct ON/OFF control~\cite{power_duration_data_driven,power_duration_experimental,power_duration_novel_ON_OFF}. However, these approaches are tailored to direct load control and do not directly characterize TCL fleets controlled through thermostat setpoint changes.
% Related power-duration and flexibility-envelope representations have also been explored in broader building and prosumer flexibility settings~\cite{power_duration_data_driven,power_duration_experimental,power_duration_novel_ON_OFF}.

Prior work on setpoint-controlled TCL fleets has primarily focused on aggregate modeling and control rather than \emph{ex-ante} flexibility characterization. PDE-based models describe the evolution of the population temperature distribution and corresponding aggregate power under setpoint changes~\cite{30_OG_PDE_Paper,callaway2009tapping,19_bashash2012_PDE_setpoint_model,6_zheng2020_PDE,21_julien2024randomized_PDE_Markov}, and related work has derived approximate transfer-function models from the underlying temperature distribution~\cite{20_kundu_setpoint_change_model_LQR}. 
% These approaches are valuable for understanding and controlling aggregate TCL response, but they do not directly quantify, \emph{ex ante}, the magnitude of an aggregate power deviation that can be delivered and the length of time for which it can be sustained. As a result, they do not provide a characterization that grid operators can use to prescribe the capacity and duration of flexible resources.
These approaches are valuable for understanding and controlling aggregate TCL response, but they do not provide grid operators with an \emph{ex-ante} characterization that can be used to prescribe the capacity and duration of flexible resources.
% These approaches are valuable for understanding and controlling aggregate TCL response, but they do not provide a directly usable \emph{ex-ante} flexibility characterization. % maybe something like: our goal is to determine the  power-duration curves for aggregated TCL fleets, which can be directly used by grid operators to prescribe capacity and duration of flexible resources??

An important consideration in this context is that aggregate TCL models are built on underlying models of individual thermal dynamics. A common representation of TCL thermal dynamics is the equivalent thermal parameter (ETP) model. First-order ETP models capture only indoor air temperature, whereas second-order ETP models also capture building thermal-mass dynamics. These additional dynamics can significantly affect the transient aggregate power response following a setpoint change~\cite{8_zhang_2d_Markov,16_liu2015_2d_markov_model,pnnl_validation,WANG2022118124}. As a result, flexibility characterizations based only on first-order dynamics may misrepresent the actual DR capability of a setpoint-controlled TCL fleet. Higher-order aggregate models have been developed using Markov chain formulations~\cite{8_zhang_2d_Markov,16_liu2015_2d_markov_model}, but an \emph{ex-ante} flexibility characterization based on such models has not been developed.
% This paper addresses this gap by characterizing the flexibility of a TCL fleet under thermostat setpoint control while accounting for TCL population dynamics. We introduce the \emph{reach-and-hold} set, which describes fleet flexibility in terms of the aggregate power deviations a TCL fleet can \emph{reach} and the durations for which those deviations can be \emph{held}. 

This paper addresses this gap by developing an \emph{ex-ante} power-duration characterization of a TCL fleet under thermostat setpoint control while accounting for TCL population dynamics. To do so, we introduce the \emph{reach-and-hold} set as a representation of fleet capability that describes the aggregate power deviations a TCL fleet can \emph{reach} and the durations for which those deviations can be \emph{held}. The reach-and-hold set thus defines the corresponding power-duration characterization of the TCL fleet. To compute this set, we use a Markov-chain aggregate model~\cite{8_zhang_2d_Markov} and formulate an optimization problem that characterizes the boundary of the reach-and-hold set for a fixed setpoint change. Because the resulting formulation becomes intractable at the necessary state bin resolutions needed for accurate transient modeling, we introduce an open-loop control policy, consistent with the operation of many current DR programs, together with an equivalent transformation that yields a tractable linear program (LP). The resulting LP represents an inner approximation of the reach-and-hold set and produces corresponding open-loop control schedules.

%The proposed model-based characterization of the reach-and-hold set and its resulting open loop temperature set-point control schedules are evaluated using realistic agent-based simulations. In addition, because the formulation depends on estimated initial temperature distributions and TCL parameters, we assess the sensitivity of both the reach-and-hold characterization and the corresponding schedules to uncertainty.
% The proposed model-based characterization of the reach-and-hold set and its resulting open-loop temperature set-point control schedules are evaluated using realistic agent-based simulations. To assess the robustness of the proposed formulation against common parametric uncertainties, we assess the sensitivity of both the reach-and-hold characterization and the corresponding schedules to variations in estimated initial temperature distributions and TCL parameters.
The proposed reach-and-hold characterization and resulting open-loop setpoint control schedules are evaluated using agent-based simulations. We also assess sensitivity to uncertainty in initial temperature distributions and TCL parameters.

% The main contributions of this paper are:
% \begin{enumerate}
%     \item introduction of the reach-and-hold set as an \emph{ex-ante} novel characterization of the capability of TCL aggregations under thermostat setpoint control.
%     \item We characterize the entire reach-and-hold set as the solution to a very large-scale linear program that embeds the Markov-chain aggregate model but only needs to compute its boundary to engender any open-loop control schedule.
%     \item  Develop a practical, but nonlinear inner approximation of the reach-and-hold set and prove equivalence to a tractable LP formulation.
%     \item We validate the resulting inner approximation and open-loop schedules with agent-based simulations and demonstrate robustness to uncertainty in initial conditions and TCL parameters.
% \end{enumerate}}
The main contributions of this paper are:
\begin{enumerate}
    \item \textbf{Introduce} the novel concept of the reach-and-hold set as an \emph{ex-ante} characterization of the demand response capability of setpoint-controlled TCL aggregations. This concept defines the entire operational reach-and-hold space rather than a single power-duration curve and inherently maps to the required control schedules.
    
    \item \textbf{Formulate} the exact boundary of this set as a large-scale LP that embeds a Markov-chain model, and then {prove} convexity in power to establish that all feasible control schedules can be derived directly from its reach-and-hold boundary.

    \item \textbf{Derive} an implementable inner approximation based on a proportional control policy, and {prove} its exact equivalence to a computationally tractable LP that efficiently engenders open-loop control schedules.

    \item \textbf{Validate} the characterized set and control schedules using agent-based simulations and assess robustness to uncertainties in initial temperature distributions and TCL parameters.
\end{enumerate}
Section~\ref{sec: TCL Modeling} presents the second-order Markov-chain aggregate model used to capture TCL fleet dynamics. Section~\ref{sec: RnH} introduces the reach-and-hold set and formulates the corresponding LP. Section~\ref{sec:RnH_Inner} presents the inner approximation as a tractable LP. Section~\ref{sec: Simulation Results} presents simulation-based analysis, and Section~\ref{sec:Conc} concludes the paper.

% The rest of the paper is organized as follows: Section~\ref{sec: TCL Modeling} presents the second-order Markov-chain aggregate model used to capture TCL fleet dynamics. Section~\ref{sec: RnH} introduces the reach-and-hold set and formulates the corresponding LP. Section~\ref{sec:RnH_Inner} presents the inner approximation as a tractable LP. Section~\ref{sec: RnH_Characterization} uses this formulation to characterize the reach-and-hold set and discuss the effects of peak limiting constraints and parameter heterogeneity. Section~\ref{sec: Simulation Results} presents simulation-based analysis, while Section~\ref{sec:Conc} concludes the paper.

\section{Modeling Thermostatically Controlled Loads}\label{sec: TCL Modeling}
Consider a TCL that operates by switching ON or OFF to maintain indoor air temperature within a predefined deadband $\Delta T_i$ around a setpoint, i.e., $[T_{\text{set},i}-\frac{\Delta T_i}{2},T_{\text{set},i}+\frac{\Delta T_i}{2}]$, where $T_{\text{set},i}$ is the thermostat setpoint temperature for device $i$. %and the subscript $i$ is used to denote the device index in the TCL population.---> VERBOSE.

In this work, we consider the specific case of an HVAC (air conditioner) as the TCL of interest; however, the work presented can be generalized to other TCLs as well. 
\subsection{Individual HVAC Model}
The dynamics of individual HVAC systems can be characterized with a second-order, single-zone ETP model~\cite{ETP_OG}:
\begin{subequations} \label{eq:ETP_2D}
    \begin{align}
          C_{\text{a},i}\dot{T}_{\text{a},i}(t) &= H_{\text{m},i}T_{\text{m},i}(t) - (U_{\text{a},i} + H_{\text{m},i})T_{\text{a},i}(t)  \nonumber \\ &\qquad\qquad  +U_{\text{a},i} T_\text{amb}(t) + Q_{\text{a},i}(t) \label{eqn:ETP2D_Ta} \\
          C_{\text{m},i}\dot{T}_{\text{m},i}(t)&=H_{\text{m},i}\left(T_{\text{a},i}(t) - T_{\text{m},i}(t)\right) + Q_{\text{m},i}(t)\label{eqn:ETP2D_Tm},
    \end{align}
\end{subequations}

\noindent where $T_{\text{a},i}(t)$ and $T_{\text{m},i}(t)$ are the air and building mass temperatures respectively. The parameters $C_{\text{a},i}$ and $C_{\text{m},i}$ represent the thermal capacitances of indoor air and building mass, while $U_{\text{a},i}$ and $H_{\text{m},i}$ are conductances of the building envelope and between inner air and solid mass, respectively. Finally, $Q_{\text{a},i}(t)$ is the net heat gain from the HVAC, $Q_{\text{m},i}(t)$ is the heat flux to the interior solid mass, and $T_\text{amb}(t)$ is the outdoor temperature. The heat gains $Q_{\text{a},i}$ and $Q_{\text{m},i}$ include contributions from internal loads, solar irradiance, and the heating or cooling provided by the HVAC system. 

The internal and solar heat gains are neglected in this work for simplicity, although they could be incorporated into the transition matrix, if they are modeled or measured. 

Herein, we consider the common ON/OFF switching logic:%~\cite{4_HeHao_2014aggregate} is adopted: 

\begin{align}\label{eqn:switching logic}
            Q_{\text{a},i}(t^+) &= \begin{cases} 
            0 & \text{if } T_{\text{a},i}(t) \leq T_{\text{set},i} - \frac{\Delta T_i}{2} \\
            \eta_i P_{\text{rated},i} & \text{if } T_{\text{a},i}(t) \geq T_{\text{set},i} + \frac{\Delta T_i}{2} \\
            Q_{\text{a},i}(t) & \text{otherwise} ,
   \end{cases}
\end{align}
where $P_{\text{rated},i}$ and $\eta_i$ are the rated power and the coefficient of performance of unit $i$, respectively. 

\subsection{Aggregate Modeling of HVAC Fleet}
The ETP model in \eqref{eq:ETP_2D} and \eqref{eqn:switching logic} is a hybrid dynamical system. To obtain a tractable aggregate representation, several models have been proposed in the literature~\cite{30_OG_PDE_Paper,8_zhang_2d_Markov,28_soudjani_formal_abstractions}. In this work, we build on the second-order Markov chain-based aggregate model in~\cite{16_liu2015_2d_markov_model,8_zhang_2d_Markov}, and modify them to include an explicit setpoint-change input. Furthermore, we adopt a single-cluster representation for the aggregate TCL model, where a \emph{cluster} denotes a group of heterogeneous TCLs represented by a common thermal/electrical parameter vector. Thus, the single-cluster model does not imply a homogeneous fleet; rather, it approximates the aggregate population using one representative parameter vector $(U_\text{a},C_\text{a},C_\text{m},H_\text{m},\eta,P_\text{rated})$. For simplicity, we also assume homogeneous setpoints and deadbands, i.e., $T_{\text{set},i}=T_{\text{set}}$ and $\Delta T_i=\Delta T$ for all devices. More generally, the population can be partitioned into groups according to both parameter vector and setpoint/deadband, and the overall aggregate response is obtained by combining the corresponding group-level models~\cite{8_zhang_2d_Markov}. This multi-group extension is discussed later in Remark~\ref{remark:heterogeneous considereations}.

Markov chain-based aggregate TCL models generally involve discretizing the temperature operating range $[T_\text{min}, T_\text{max}]$\footnote{Since we are interested in setpoint control, we assume that the temperature deadband $[T_\text{set}-\frac{\Delta T}{2},T_\text{set}+\frac{\Delta T}{2}] \subset [T_{\min}, T_{\max}]$.} into discrete temperature bins and the transition rates from/to each of the temperature bins are calculated or estimated~\cite{24_mathieu_state_nodate,8_zhang_2d_Markov}. To construct the aggregate Markov model, the air and mass temperatures are discretized into $N_\text{a}$ and $N_\text{m}$ bins, respectively, yielding a total of $N := N_\text{a}N_\text{m}$ unique air--mass temperature pairs. Furthermore, since each temperature bin can correspond to either a device being in ON or OFF state, a total of $2N$ bins are needed. Let $x[k]\in\mathbb{R}^{2N}$ denote the population state vector, whose entries give the fraction of devices in each temperature bin at time step $k$. The vector $x[k]$ is partitioned as follows:
\begin{equation}\label{eqn:x_ON_OFF_partion}
    x[k]=\begin{bmatrix}
        x_\text{on}[k]\\
        x_\text{off}[k]
    \end{bmatrix}
\end{equation}
where $x_\text{on}[k]$ and $x_\text{off}[k]\in\mathbb{R}^N$ correspond to the ON and OFF distribution vectors, respectively. 

\begin{figure}[!t]
    \centering
    \includegraphics[width=0.95\linewidth]{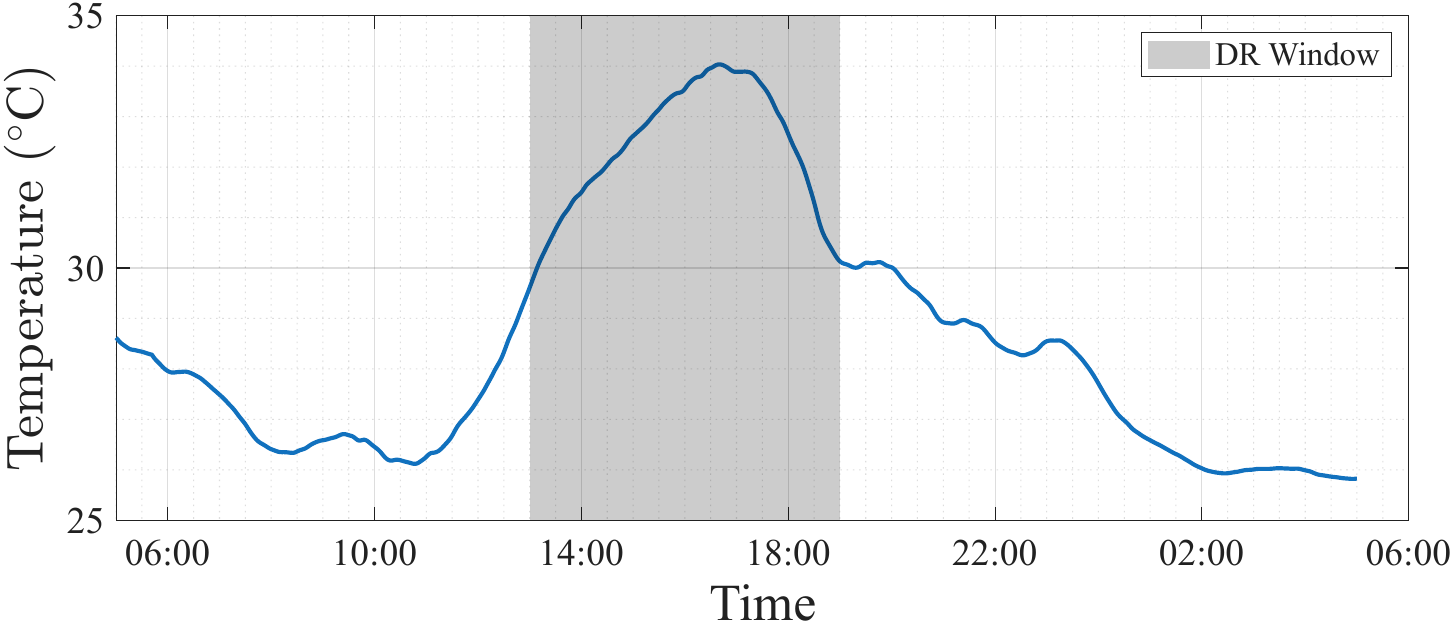}
    \caption{Outdoor temperature profile with shaded region showing considered DR time window. }
    \label{fig:T_amb}
\end{figure}

The evolution of the state vector $x[k]$ for a fixed temperature setpoint can be expressed by a discrete linear time-varying (LTV) system~\cite{16_liu2015_2d_markov_model}, given by 
\begin{subequations}\label{eqn: Markov_model_LTV} \begin{align}
x[k+1] &= A[k]x[k]\\
    x[0] &= x_0,
\end{align}
\end{subequations}
where $x_0$ is the initial temperature distribution. The state transition probabilities are calculated or estimated from data and populated in the state transition matrix $A[k]\in\mathbb{R}^{2N\times2N}$ \cite{8_zhang_2d_Markov,16_liu2015_2d_markov_model}. The matrix $A[k]$ is a function of $T_\text{set}$, $\Delta T$ and $T_\text{amb}[k]$ which are time-varying and hence the LTV nature of~\eqref{eqn: Markov_model_LTV}. For the case studies considered in this paper, $A[k]$ is computed using the $T_\text{amb}$ trajectory shown in Fig.~\ref{fig:T_amb}; however, the proposed framework can be applied to other ambient temperature trajectories as well.

The aggregate power consumption, $P_\text{agg}[k]$, of the fleet is obtained by considering the total fraction of ON devices as shown below
\begin{align}\label{eqn:Markov_Pagg}
    P_\text{agg}[k]= C_\text{ON}x[k] ,
\end{align}
where $C_\text{ON}:=N_\text{TCL}P_\text{rated}[\mathbf{1}^\top_N~ \mathbf{0}_N^\top]$ and $N_\text{TCL}$ is the number of TCL devices in the fleet. Note that if a multi-cluster representation is used, each cluster is modeled analogously with its own state vector and transition matrix. The total aggregate power is obtained by summing the cluster-level aggregate powers. Next, we describe how setpoint changes are incorporated explicitly into the aggregate model.

\subsection{Setpoint Control}
The model in~\eqref{eqn: Markov_model_LTV} does not include an explicit input for setpoint changes. We therefore modify it to include such an input, which will later be optimized to characterize the feasible aggregate power shifts of the TCL fleet. Specifically, we consider a transition from an initial setpoint $T_{\text{set,n}}$ to a new setpoint $T_{\text{set,dr}}$. The new setpoint temperature is applied to a subset of the population; therefore, devices that undergo setpoint change will have different underlying dynamics governing their temperature evolution. Hence, we distinguish between two device groups: those that retain their initial setpoint $T_{\text{set,n}}$ and those that undergo a setpoint change to $T_{\text{set,dr}}$. Their corresponding temperature distribution vectors are denoted by $x_\text{n}[k]$ and $x_\text{dr}[k]\in \mathbb{R}^{2N}$, respectively. Because the state transition matrix depends on the setpoint temperature, we define $A_\text{n}[k]$ and $A_\text{dr}[k]\in \mathbb{R}^{2N\times2N}$ as the transition matrices associated with $T_{\text{set,n}}$ and $T_{\text{set,dr}}$, respectively. Lastly, we introduce the system input $u[k]\in\mathbb{R}^{2N}$ where each element represents the proportion of the population at the corresponding temperature bin that change setpoint temperature at timestep $k$. With these definitions, the aggregate model in~\eqref{eqn: Markov_model_LTV} is extended to incorporate a fixed setpoint-change policy as follows:

\begin{subequations} \label{eqn:markov_with_setpoint_u[k]}
     \begin{align}
         x_\text{n}[k+1]&=A_\text{n}[k](x_\text{n}[k]-u[k]),\label{eqn: x_c_update}\\
         x_\text{dr}[k+1]&=A_\text{dr}[k](x_\text{dr}[k]+u[k]),\label{eqn: x_h_update}\\
         x_\text{n}[0]=& x_{\text{n,}0}, \quad x_\text{dr}[0]= x_{\text{dr,}0}\label{eqn: x_IC}\\
          P_\text{agg}[k]&= C_\text{ON}(x_\text{n}[k] + x_\text{dr}[k])\label{eqn:Pagg_with_setpoint}
     \end{align}
 \end{subequations}
 \noindent where $x_{\text{n,}0}$ and $x_{\text{dr,}0}$ are the initial temperature distributions for each of the $x_\text{n}[k]$ and $x_\text{dr}[k]$ populations respectively. 
 
\begin{figure}[!t]
    \centering
    \includegraphics[width=0.95\linewidth]{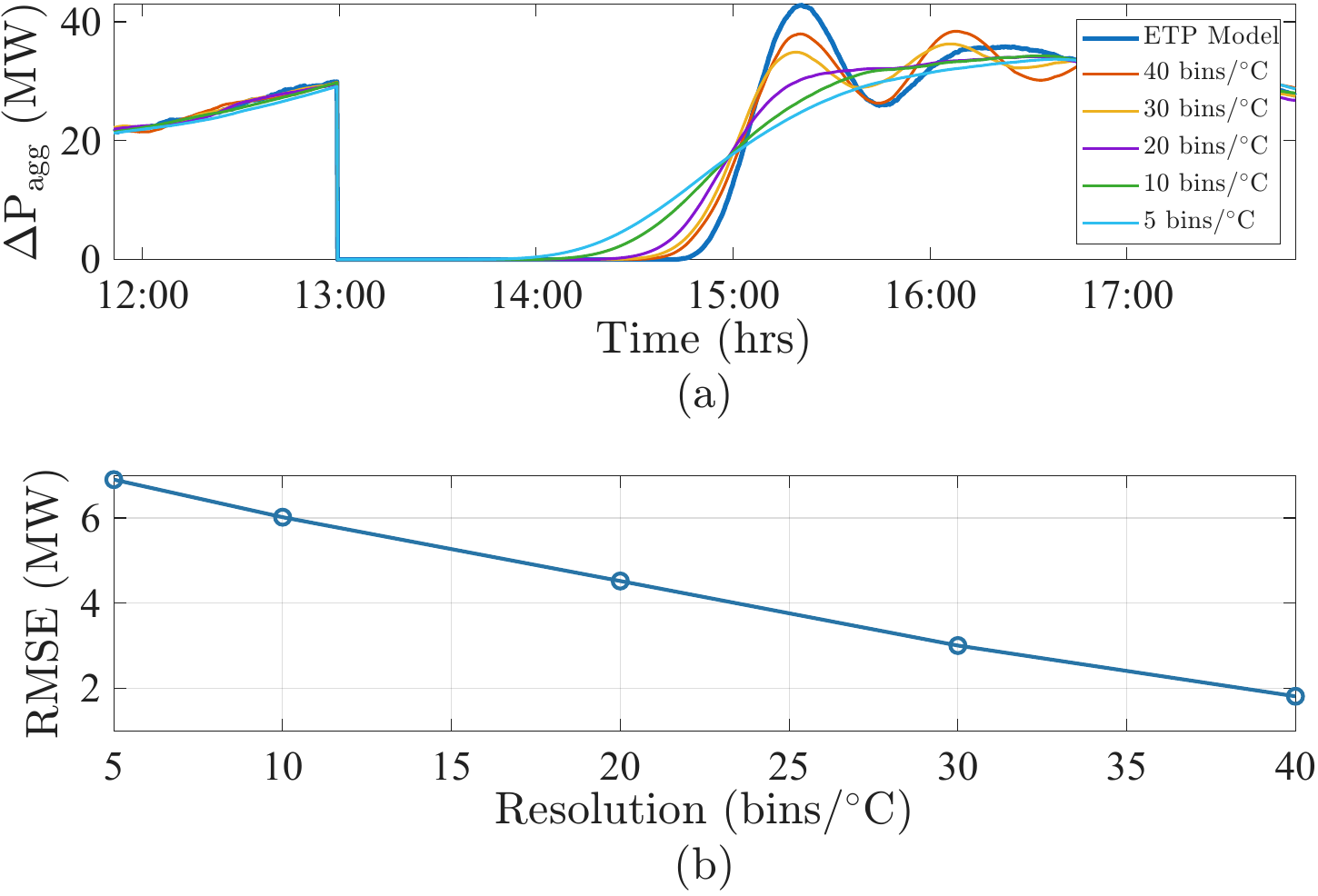}
    \caption{(a) Aggregate HVAC power consumption with setpoint change from 20$^\circ$C to 22$^\circ$C at $t$=13:00 obtained from agent-based simulation (ETP model) and Markov models with different temperature discretization resolutions. (b) MSE associated with each Markov model's power characterization.}
    \label{fig:Markov_resolution_comparison_with_setpoint}
\end{figure}

 Practically, the number of devices in a temperature bin that change setpoint cannot exceed the number of devices present in that bin. Thus, for each time, $k$, we must have that $u[k] \in \mathcal{X}_{u}$, where $\mathcal{X}_{u}$ is the set of admissible inputs:
\begin{align}\label{eqn:Admissible_input}
\mathcal{X}_{u}(x_\text{n}[k]) := \left\{ u: \mathbf{0}_{2N}\preceq u\preceq x_\text{n}[k]\right\}. 
 \end{align}
 The accuracy of the aggregate model in~\eqref{eqn:markov_with_setpoint_u[k]} depends on the chosen bin resolution. To illustrate this, the same setpoint change, from 20$^\circ$C to 22$^\circ$C, is applied to the fleet at $t=13{:}00$, and the resulting aggregate power trajectories from Markov models with different bin resolutions are compared in Fig.~\ref{fig:Markov_resolution_comparison_with_setpoint}. In addition, the mean squared error (MSE) between each Markov model and the agent-based ETP simulation during the transient event is computed and shown in Fig.~\ref{fig:Markov_resolution_comparison_with_setpoint}.
Clearly, the bin resolution has a significant effect on model error during transients. Therefore, in this paper, we leverage a higher bin resolution (40 bins$/^\circ$C) to avoid inaccurate and overly conservative flexibility characterizations for DR. 
Now that we have defined and validated the aggregate TCL model, we use it next to introduce the \emph{reach-and-hold} set.

\section{The Reach and Hold Set}\label{sec: RnH}

\subsection{Motivation behind the reach-and-hold set}
\begin{figure}[!t]
    \centering
    \includegraphics[width=0.95\linewidth]{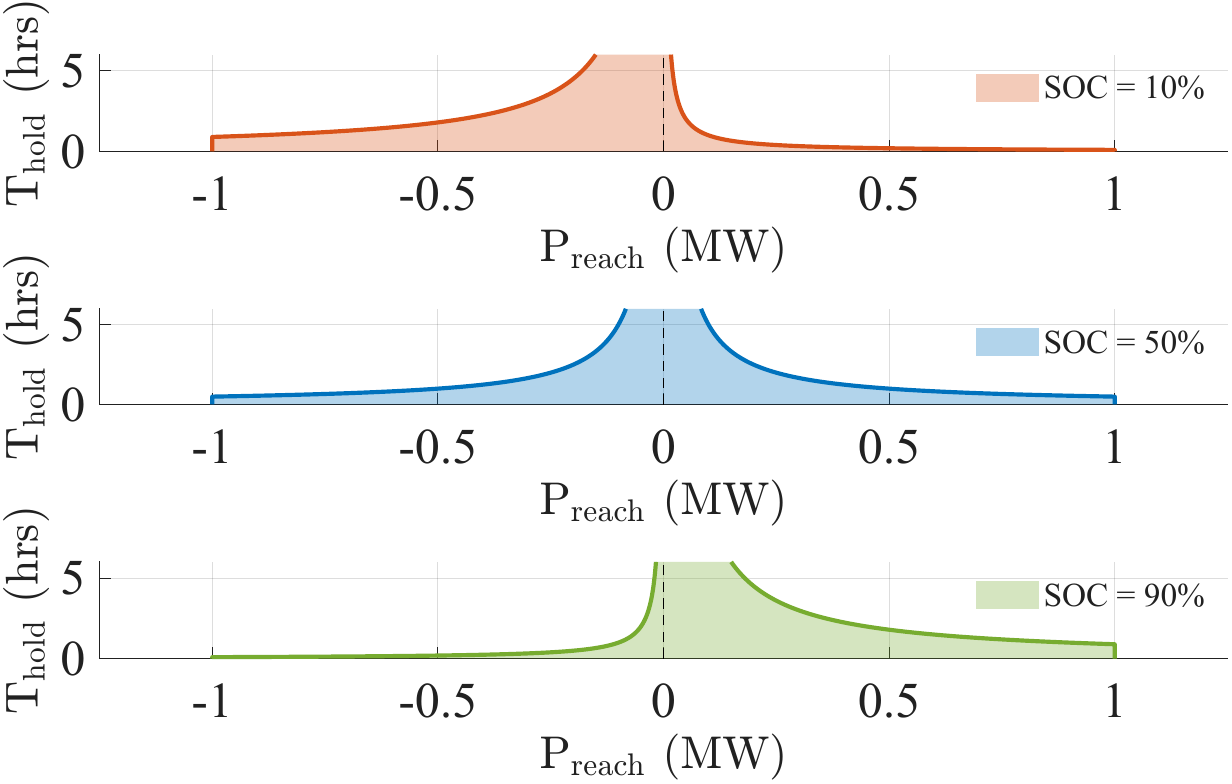}
    \caption{Reach-and-hold sets for a 1C battery with initial SOC of 10\%, 50\% and 90\%.}
    \label{fig:Battery_Rnh}
\end{figure}

 The reach-and-hold set characterizes the changes in power ($P_\text{reach}$) that a resource can \emph{reach} (i.e., change) and the duration for which the power change can be \emph{held} (i.e., sustained). This set can be defined for different resources, e.g.,  batteries, generators, or fleets of TCLs. For the case of a battery, for example, this set is straightforward to define based on its power limits, energy limits and initial state of charge (SOC), see Fig.~\ref{fig:Battery_Rnh} for the reach-and-hold set of a 1MW/1MWh battery at 10\%, 50\% and 90\% initial SOC.  As illustrated in Fig.~\ref{fig:Battery_Rnh}, a high initial SOC allows for prolonged discharging ($P_\text{reach}\ge0$) but limits charging capability ($P_\text{reach}\le0$), and vice versa. The boundaries of the set are defined by the battery's power and energy limits.

This framework can be extended to aggregations of TCLs. By changing the devices' setpoint, the fleet's aggregate power consumption can be modulated as shown in Fig.~\ref{fig:RnH_event}. In the case of HVACs, increasing the setpoint temperature lowers aggregate power consumption, analogous to discharging a battery, while increasing power consumption is analogous to charging a battery. Unlike a battery; however, a TCL fleet exhibits thermal dynamics and a non-zero nominal power consumption, necessitating that $P_\text{reach}$ be measured with respect to the fleet's time-varying nominal power consumption as illustrated in Fig.~\ref{fig:RnH_event}. Furthermore, unlike the SOC of a simple battery, the ``state'' of a TCL fleet is dictated by the fleet's temperature distribution, which is high dimensional and is a function of time-varying ambient conditions. These added complexities make characterizing the TCLs' reach-and-hold challenging. 

\subsection{Defining the reach-and-hold Set}
We begin by introducing the discrete horizon $\mathcal{K}:=\{0,\ldots, K\}$, over which the system dynamics will be simulated in order to compute the {\it reach-and-hold} sets. Here, $K > 0$ is chosen such that $\mathcal{K}$ spans the entire DR event. We denote the DR event window by $\mathcal{K}_\text{hold}:=\{k_0, \dots , k_0+\ceil{\frac{T_\text{hold}}{\Delta t}}\}\subseteq \mathcal{K}$, where ${k_0\ge0}$ is the timestep corresponding to the start of the DR event and $\Delta t$ is the step size of the discretization. As an example, in Fig.~\ref{fig:RnH_event}, $K$ is the total number of timesteps considered throughout the day and $k_0$ is the timestep corresponding to 13:00. Thus, $\mathcal{K}_\text{hold}$ defines the interval within $\mathcal{K}$ during which the fleet is required to sustain a desired power deviation.

Additionally, we define the nominal power consumption trajectory $P_\text{nom}[k]$ corresponding to the power consumption with no setpoint change. Given a forecast of $T_\text{amb}[k]$, $P_\text{nom}[k]$ can be computed using~\eqref{eqn:markov_with_setpoint_u[k]} by setting $u[k]=0~\forall k \in \mathcal{K}$.

Next, we define the {\it change} in aggregate power $\Delta P_\text{agg}[k]$:

\begin{figure}[!t]
    \centering
    \includegraphics[width=0.95\linewidth]{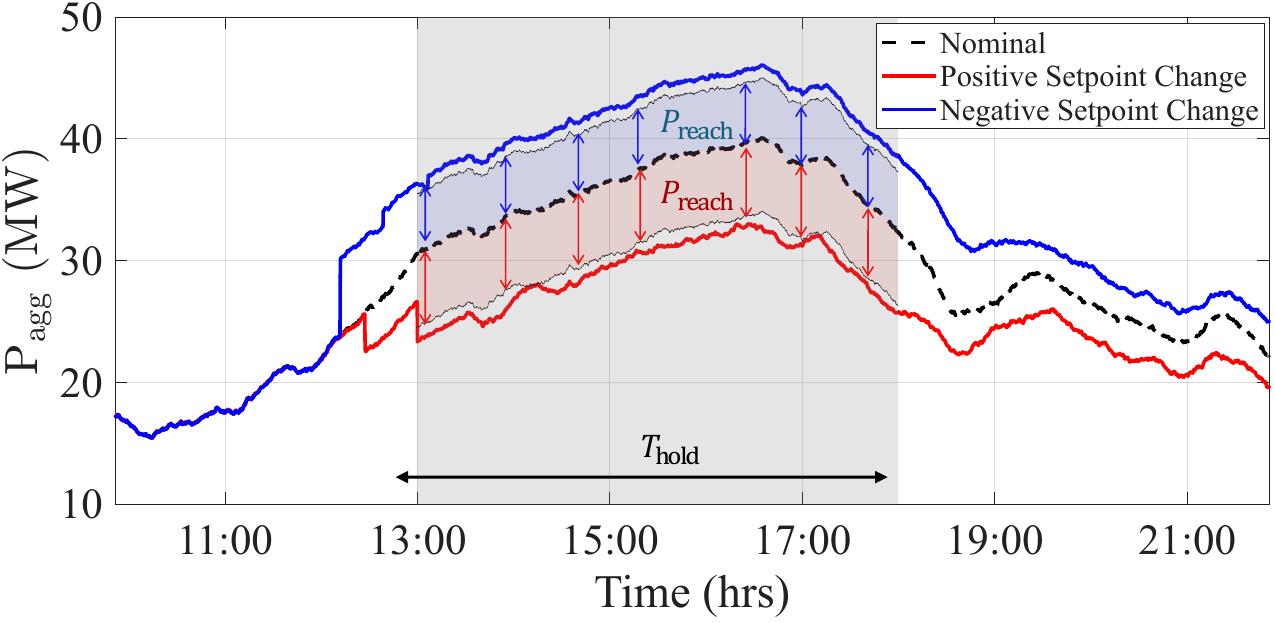}
    \caption{Change in aggregate power consumption $P_\text{agg}$ during a DR event with corresponding $P_\text{reach}$ and $T_\text{hold}$ values.}
    \label{fig:RnH_event}
\end{figure}

\begin{align}
    \Delta P_\text{agg}[k]:=P_\text{nom}[k] - P_\text{agg}[k].
\end{align}

With these notations, we formally define the reach-and-hold set of a TCL fleet as follows:

\begin{definition}[Reach and Hold Set]\label{def:reach-hold}

Let $k_0$ be the first timestep of the DR event of interest. The \emph{reach–and–hold set} at timestep $k_0$, denoted by $\mathcal{R}$, is the set of all pairs $(P_\text{reach}, T_\text{hold})$
for which there exists an admissible control input $u[\cdot] \in \mathcal{X}_u$ such that $\forall k \in \mathcal{K}_\text{hold}$:
\[
\begin{cases}
0\le P_\text{reach} \le \Delta P_\text{agg}[k], & \text{if } \Delta P_\text{agg}[k] \ge 0\\[3pt]
\Delta P_\text{agg}[k] \le P_\text{reach} \le 0, & \text{otherwise},
\end{cases}
\]
where $\mathcal{K}_\text{hold} := \{k_0, \ldots, k_0 + \lceil T_\text{hold}/\Delta t \rceil\}$ is the DR event window.
\end{definition}
In words, $(P_\text{reach},T_\text{hold})\in\mathcal{R}$ means the TCL fleet can be
driven, under dynamics~\eqref{eqn:markov_with_setpoint_u[k]}, to sustain at least $P_\text{reach}$
aggregate deviation from nominal for $\lceil T_\text{hold}/\Delta t \rceil + 1$ consecutive timesteps.   

By this definition, $\Delta P_\text{agg}[k]$ is allowed to overshoot the value of $P_\text{reach}$ (as also illustrated in Fig.~\ref{fig:RnH_event}). Methods to limit the size of this overshoot are discussed later in section~\ref{sec:RnH_Inner}.

Finding the reach-and-hold set does not require an exhaustive search of all feasible $(P_\text{reach},T_\text{hold})$ pairs. Instead, we only need to find the boundary of $\mathcal{R}$ ($\mathrm{bd}(\mathcal{R})$), which is the maximum (minimum) $P_\text{reach}$ for a given $T_\text{hold}$ under a fixed positive (negative) setpoint change control policy. This is thanks to the following convexity property of $\mathcal{R}$.

\begin{prop}\label{prop1} 

 If $({P}_\text{reach},\tau) \in \mathrm{bd}(\mathcal{R})$ then $  ({P},\tau) \in \mathcal{R} \,\, \forall  P \in [\min\{0,P_\text{reach}\},\max\{0,P_\text{reach}\}]$.
\end{prop}

\begin{proof}
    Consider the case of ${P}_\text{reach} \ge 0$. If $({P}_\text{reach},\tau) \in \mathrm{bd}(\mathcal{R})$, then there exists a corresponding input trajectory $\hat{u}[k]\in\mathcal{X}_u$ (defined in~\eqref{eqn:Admissible_input}). Since $\hat{u}[k]\in\mathcal{X}_u$ then $\theta \hat{u}[k]\in \mathcal{X}_u$ for $\theta\in[0,1]$. Now, it can be shown that $\Delta P_\text{agg}[k]$ has a linear relationship with $u[k]$. Thus, $\theta  \hat{u}[k]$ begets output deviation 
    $\theta \hat{P}_\text{reach}=:P$.
    Therefore, $(P,\tau)\in \mathcal{R}$ \,\,$\forall P\in[0,P_\text{reach}]$. 
    The case of ${P}_\text{reach} \le 0$ is similar.
\end{proof}

Proposition~\ref{prop1} shows that, for a fixed $T_\text{hold}$ value, the reach-and-hold set is convex in $P_\text{reach}$. 
Next, we formulate the optimization problems to find the boundary of the reach-and-hold set for both positive and negative setpoint changes. We make the following assumptions and later discuss their implications in Section~\ref{sec: Simulation Results}:
\begin{itemize}
    \item \textbf{Assumption 1:} The initial temperature distributions of the TCL fleet ($x_{\text{n},0}$ and $x_{\text{dr},0}$) are known. 
    \item \textbf{Assumption 2:} The mean ETP parameter values ($U_\text{a},C_\text{a},C_\text{m},H_\text{m},\eta,P_\text{rated}$) of the TCL fleet are known and used to characterize the Markov-based aggregate model. 
\end{itemize}

\subsection{Computing the boundary of the reach-and-hold set}

Depending on the direction of the applied setpoint change, $\Delta \Pagg[k]$ can be either positive or negative. In most DR events, the goal is to reduce aggregate power consumption (i.e. $\Delta \Pagg[k]\ge0$). We will thus consider this case first. The case with $\Delta \Pagg[k]\le0$, corresponding to a negative setpoint change for air conditioners, will be discussed later. 

In the case where $\Delta \Pagg[k]\ge0$, the boundary of the reach-and-hold set corresponds to the maximum achievable value of $P_\text{reach}$ at each value of $T_\text{hold}$. From Definition~\ref{def:reach-hold}, $P_\text{reach}$ lower bounds $\Delta \Pagg[k]$; hence, for a given $\Delta \Pagg[k]$ and $\mathcal{K}_\text{hold}$, the largest value of $P_\text{reach}$ is bounded by the smallest value of $\Delta \Pagg[k]$, i.e.,  $P_\text{reach}=\min_{k\in\mathcal{K}_\text{hold}} \Delta \Pagg[k]$ over the DR window. Therefore, determining the boundary of the reach-and-hold set reduces to maximizing this bound on $\Delta \Pagg[k]$ over all input sequences. This naturally leads to the following max-min formulation:       

\begin{subequations}\label{eqn:Opt_intractable_u[k]}
\begin{align}
 P_\text{reach}^+=\max_{u[k]} & \min_{k\in \mathcal{K}_\text{hold}}\left\{ \Delta P_\text{agg}[k]\right\}\\
\textrm{s.t.} &\qquad \eqref{eqn: x_c_update}-\eqref{eqn:Pagg_with_setpoint} \label{eqn:opt_dynamics_constraints}\\
&\Delta P_\text{agg}[k] =P_\text{nom}[k] - P_\text{agg}[k]\quad \forall k \in \mathcal{K} \label{eqn:opt_delta_pagg_cosntraints}\\
  &\quad 0\preceq u[k]\preceq x_\text{n}[k] \quad \forall k\in \mathcal{K}\label{eqn:opt_input_constraints_u[k},
\end{align}
\end{subequations}

Note, the state transition matrices $A_\text{n}[k]$ and $A_\text{dr}[k]$ in~\eqref{eqn: x_c_update} and~\eqref{eqn: x_h_update} are chosen such that they correspond to the initial and final setpoint temperature values, respectively.

Conversely, a negative setpoint change results in $\Delta P_\text{agg}[k]\le 0$, as shown in Fig.~\ref{fig:RnH_event}. In this case, $P_\text{reach}$ upper bounds $\Delta \Pagg [k]$ and the boundary of the reach-and-hold set corresponds to the minimum achievable value of $P_\text{reach}$ at each $T_\text{hold}$. Following the same reasoning from before, finding the boundary of the reach-and-hold set can be formulated as the following min-max problem: 

\begin{subequations}\label{eqn:Opt_intractable_u[k]2}
\begin{align}
 P_\text{reach}^- = \min_{u[k]} & \max_{k\in \mathcal{K}_\text{hold}}\left\{ \Delta P_\text{agg}[k]\right\}\\
\textrm{s.t.} &\qquad \eqref{eqn:opt_dynamics_constraints}-\eqref{eqn:opt_input_constraints_u[k}.
\end{align}
\end{subequations}

Thus, for a given hold time $T_\text{hold}$, we can solve~\eqref{eqn:Opt_intractable_u[k]} and ~\eqref{eqn:Opt_intractable_u[k]2} to compute boundary points $(P_\text{reach}^+,T_\text{hold})$ and $(P_\text{reach}^-,T_\text{hold})$, which correspond to fixed positive and negative setpoint changes, respectively. These points yield the convex interval of reachable power deviations with a hold time of $T_\text{hold}$, $\mathcal{R}_{T_\text{hold}} := \{\, (P, T_\text{hold}) \in \mathbb{R}^2 \mid P \in [P_\text{reach}^-, P_\text{reach}^+] \,\} \subset \mathcal{R}$, where $0 \in \mathcal{R}_{T_\text{hold}}$. Solving ~\eqref{eqn:Opt_intractable_u[k]} and ~\eqref{eqn:Opt_intractable_u[k]2} repeatedly for desired hold times $T_\text{hold}\in [h_\text{min},h_\text{max}]$, engenders the reach-and-hold set via $\mathcal{R} = \cup_{T_\text{hold}\in [h_\text{min},h_\text{max}]} \mathcal{R}_{T_\text{hold}}$. 

As shown in Section~\ref{sec: TCL Modeling}, a high bin resolution is necessary for an accurate characterization of the fleet dynamics and, by extension, of $\mathcal{R}$. This greatly increases computational complexity of solving~\eqref{eqn:Opt_intractable_u[k]} and ~\eqref{eqn:Opt_intractable_u[k]2}, as discussed next.

\subsection{Problem Tractability}

Although the control input $u[k]$ is general enough to capture arbitrary setpoint switching policies, it dramatically increases the complexity of \eqref{eqn:Opt_intractable_u[k]} and \eqref{eqn:Opt_intractable_u[k]2} with $(6N+1)\lvert \mathcal{K}\rvert$ variables and $(8N+1)\lvert \mathcal{K}\rvert$ constraints. The scale of \eqref{eqn:Opt_intractable_u[k]} and \eqref{eqn:Opt_intractable_u[k]2} is quantified in Table~\ref{tab:bin_resolution_complexity}, which details the number of variables and constraints for a system with a 6-hour time horizon ($\mathcal{K}$) and a 5°C operating range. 

As can be seen, even a moderate bin resolution of $r=20$ bins$/\C$ begets 120~million variables and 170~million constraints, rendering the problem computationally intractable.

\begin{table}[h]
  \centering
  \caption{Problem complexity vs. bin resolutions ($r$)}
  \begin{tabular}{@{}ccc@{}}
    \toprule
    Bin Resolution $r$  & Number of  & Number of  \\
    (bins$/^\circ$C) & Variables & Constraints \\
    \midrule
%    5  & $8.1\times 10^6$  & $1.1\times 10^7$ \\
    10 & $3.2\times 10^7$ & $4.3\times 10^7$ \\
    20 & $1.2\times 10^8$ & $1.7\times 10^8$ \\
    40 & $5.2\times 10^8$ & $7.0\times 10^8$ \\
    \bottomrule
  \end{tabular}
  \label{tab:bin_resolution_complexity}
\end{table}

Beyond the computational burden, implementing the optimal input sequence $u[k]$ is also challenging, as it would require knowing each device’s temperature with impractically high precision (e.g., $r=40$ corresponds to sensor accuracy of $\pm 0.025^\circ\text{C}$)

Given these challenges, we next introduce a novel control structure for $u[\cdot]$ that (\textit{i}) is implementable on standard thermostat hardware, and (\textit{ii}) ensures a tractable optimization problem. The caveat is that it yields an {\it inner approximation} of the reach-and-hold set.

\section{Tractable Reach-and-Hold Set Formulation}\label{sec:RnH_Inner}
\subsection{Proportional Control Policy}
To make the optimization problems in~\eqref{eqn:Opt_intractable_u[k]} and~\eqref{eqn:Opt_intractable_u[k]2} computationally tractable, we restrict the control input to the following proportional control policy:
\begin{align}
    u[k] = \alpha[k] x_{\mathrm n}[k], \qquad \alpha[k]\in[0,1].
    \label{eqn:alpha_control_law_new}
\end{align}
Here, $\alpha[k]$ represents the proportion of the population in the nominal-setpoint group $x_{\mathrm n}[k]$ that changes setpoint at time $k$. From a practical standpoint, this control law can be implemented by broadcasting the scalar $\alpha[k]$ to all devices, where each device interprets $\alpha[k]$ as the probability of switching to the demand-response setpoint.

Substituting~\eqref{eqn:alpha_control_law_new} into the dynamics~\eqref{eqn:markov_with_setpoint_u[k]} gives
\begin{subequations} \label{eqn:markov_with_alpha[k]}
     \begin{align}
         x_\text{n}[k+1]&=(1-\alpha[k])A_\text{n}[k]x_\text{n}[k],\label{eqn: x_c_update_alpha}\\
         x_\text{dr}[k+1]&=A_\text{dr}[k](x_\text{dr}[k]+\alpha[k]x_\text{n}[k])\label{eqn: x_h_update_alpha}\\
         \eqref{eqn: x_IC} &\text{ and } \eqref{eqn:Pagg_with_setpoint}.
     \end{align}
 \end{subequations}

Although the control law \eqref{eqn:markov_with_alpha[k]} reduces the decision space in the optimization problems \eqref{eqn:Opt_intractable_u[k]} and \eqref{eqn:Opt_intractable_u[k]2}, the bilinear term $(1-\alpha[k])x_{\mathrm n}[k]$ in~\eqref{eqn: x_c_update_alpha} makes the problem non-convex and NP-hard. To overcome this, we next derive an equivalent representation in which aggregate power has an affine relationship with the control input allowing us to formulate the reach-and-hold characterization problem as a tractable LP.

\subsection{Equivalent Linear Reformulation}
To remove the bilinearity, we first define $s[k]$ as the fraction of the total population that remains in the nominal-setpoint group at time $k$ and is given by
\begin{align}
    s[k] := \mathbf{1}_{2N}^\top x_{\mathrm n}[k], 
    \qquad 
    s[0] = \beta := \mathbf{1}_{2N}^\top x_{\mathrm n,0}\in(0,1].
    \label{eqn:s_def}
\end{align}
Since $A_{\mathrm n}[k]$ is a stochastic transition matrix, left multiplying both sides of~\eqref{eqn: x_c_update_alpha} by $\mathbf{1}_{2N}^\top$ gives
\begin{align}
    s[k+1] = (1-\alpha[k])s[k].
    \label{eqn:s_recursion}
\end{align}

Next, we define the normalized state $\tilde x_{\mathrm n}[k]$ as
\begin{align}
    \tilde x_{\mathrm n}[k] := \frac{x_{\mathrm n}[k]}{s[k]},
    \qquad \text{for } s[k]>0,
    \label{eqn:xn_tilde_def_new}
\end{align}
so that $x_{\mathrm n}[k] = s[k]\tilde x_{\mathrm n}[k]$ and $\mathbf{1}_{2N}^\top \tilde x_{\mathrm n}[k]=1$.  
Substituting $x_{\mathrm n}[k]=s[k]\tilde x_{\mathrm n}[k]$ and~\eqref{eqn:s_recursion} into~\eqref{eqn: x_c_update_alpha} yields
\begin{align}
    \tilde x_{\mathrm n}[k+1] = A_{\mathrm n}[k]\tilde x_{\mathrm n}[k].
    \label{eqn:xn_tilde_dynamics_new}
\end{align}
Thus, the normalized state $\tilde x_{\mathrm n}[k]$ evolves autonomously and is independent of the control input. Hence, given $x_\text{n}[0]$ and a temperature forecast, $\tilde x_{\mathrm n}[k+1] $ can be pre-computed before optimization.

We now define the absolute fraction of the total population that switches setpoint at time $k$:
\begin{align}
    \tilde\alpha[k] := \alpha[k]\,s[k].
    \label{eqn:alpha_tilde_def_new}
\end{align}
Using~\eqref{eqn:alpha_tilde_def_new}, the dynamics in~\eqref{eqn:s_recursion} become the affine recursion
\begin{align}
    s[k+1] = s[k] - \tilde\alpha[k],
    \label{eqn:s_affine_new}
\end{align}
and hence can be expressed as
\begin{align}
    s[k] = \beta - \sum_{l=0}^{k-1}\tilde\alpha[l].
    \label{eqn:s_closed_form_new}
\end{align}

Using~\eqref{eqn:alpha_tilde_def_new} and~\eqref{eqn:xn_tilde_def_new}, the DR-group dynamics~\eqref{eqn: x_h_update_alpha} can be rewritten as
\begin{align}
    x_{\mathrm{dr}}[k+1] 
    = A_{\mathrm{dr}}[k]\Bigl(x_{\mathrm{dr}}[k]
    + \tilde\alpha[k]\tilde x_{\mathrm n}[k]\Bigr).
    \label{eqn:xdr_linear_new}
\end{align}
The aggregate power can then be written as
\begin{align}
    P_{\mathrm{agg}}[k] = C_{\mathrm{ON}}\bigl(x_{\mathrm{dr}}[k] + s[k]\tilde x_{\mathrm n}[k]\bigr).
    \label{eqn:pagg_linear_new}
\end{align}

Collecting~\eqref{eqn:xn_tilde_dynamics_new}, \eqref{eqn:s_affine_new}, \eqref{eqn:xdr_linear_new}, and~\eqref{eqn:pagg_linear_new}, we obtain the following equivalent linear representation:
\begin{subequations}\label{eqn:equivalent_linear_system_new}
\begin{align}
    \tilde x_{\mathrm n}[k+1] &= A_{\mathrm n}[k]\tilde x_{\mathrm n}[k] \label{eqn:eq_lin_a}\\
    s[k+1] &= s[k]-\tilde\alpha[k],\label{eqn:eq_lin_b}\\
    x_{\mathrm{dr}}[k+1] &= A_{\mathrm{dr}}[k]\Bigl(x_{\mathrm{dr}}[k]+ \tilde\alpha[k]\tilde x_{\mathrm n}[k]\Bigr) \label{eqn:eq_lin_c}\\
    P_{\mathrm{agg}}[k] &= C_{\mathrm{ON}}\bigl(x_{\mathrm{dr}}[k] + s[k]\tilde x_{\mathrm n}[k]\bigr). 
    \label{eqn:eq_lin_d}
\end{align}
\end{subequations}

The formulation in~\eqref{eqn:equivalent_linear_system_new} is equivalent to~\eqref{eqn:markov_with_alpha[k]}, but has the important advantage that $\tilde{x}_{\mathrm n}[k]$ can be pre-computed offline for a given $x_{\mathrm n,0}$ and forecasted ambient temperature. As a result, for given initial conditions, the remaining variables $s[k]$, $x_{\mathrm{dr}}[k]$, and $P_{\mathrm{agg}}[k]$ can all be expressed as affine functions of the input sequence $\tilde{\alpha}[\cdot]$. This allows the state variables to be eliminated from the optimization problem, yielding a reduced formulation in which the scalar input $\tilde{\alpha}[k]$ is the only decision variable at each time step.

For given initial conditions $x_{\mathrm n,0}$ and $x_{\mathrm{dr},0}$, the state $x_{\mathrm{dr}}[k]$ can be expressed as
\begin{align} 
x_{\mathrm{dr}}[k] = \Phi_{\mathrm{dr}}[k,0]x_{\mathrm{dr},0} + \sum_{l=0}^{k-1} \Phi_{\mathrm{dr}}[k,l]\tilde{x}_{\mathrm n}[l]\, \tilde{\alpha}[l],
\label{eqn:xdr_state_transition}
\end{align}
where $\Phi_{\mathrm{dr}}[k,l]$ is the state transition matrix associated with $A_{\mathrm{dr}}[k]$ and is given by
\begin{align}
    \Phi_{\mathrm{dr}}[k,l]
    :=
    \begin{cases}
        A_{\mathrm{dr}}[k-1]A_{\mathrm{dr}}[k-2]\cdots A_{\mathrm{dr}}[l], & k>l\ge 0,\\
        I, & k=l.
    \end{cases}
\end{align}

The nominal power trajectory $P_{\mathrm{nom}}[k]$ is defined as the aggregate power trajectory with no setpoint change applied. Under the proportional policy~\eqref{eqn:alpha_control_law_new}, this corresponds to $\alpha[k]=0$, and hence to $\tilde{\alpha}[k]=0$. In that case, \eqref{eqn:s_closed_form_new} gives $s[k]=\beta$, and \eqref{eqn:xdr_state_transition} reduces to
\begin{align}
    x_{\mathrm{dr}}[k] = \Phi_{\mathrm{dr}}[k,0]x_{\mathrm{dr},0}.
\end{align}
Substituting these expressions into~\eqref{eqn:eq_lin_d} yields
\begin{align}
    P_{\mathrm{nom}}[k]
    &=
    \beta\,C_{\mathrm{ON}}\tilde{x}_{\mathrm n}[k]
    + C_{\mathrm{ON}} \Phi_{\mathrm{dr}}[k,0]x_{\mathrm{dr},0}.
    \label{eqn:Pnom_new}
\end{align}
For \(k\in\mathcal{K}\setminus\{0\}\), substituting~\eqref{eqn:s_closed_form_new} and~\eqref{eqn:xdr_state_transition} into~\eqref{eqn:eq_lin_d} gives
\begin{align}
    P_{\mathrm{agg}}[k] = P_{\mathrm{nom}}[k] - \sum_{l=0}^{k-1} h_{k,l}\,\tilde{\alpha}[l],
    \label{eqn:Pagg_affine_new}
\end{align}
where
\begin{align}
    h_{k,l}
    &:=
    C_{\mathrm{ON}}\tilde{x}_{\mathrm n}[k]
    -
    C_{\mathrm{ON}}\Phi_{\mathrm{dr}}[k,l]\tilde{x}_{\mathrm n}[l].
    \label{eqn:h_coeff_new}
\end{align}
Therefore, $\Delta P_{\mathrm{agg}}[k]$ can be expressed as
\begin{align}
    \Delta P_{\mathrm{agg}}[k]
    :=
    P_{\mathrm{nom}}[k]-P_{\mathrm{agg}}[k]
    =
    \sum_{l=0}^{k-1} h_{k,l}\,\tilde{\alpha}[l].
    \label{eqn:DeltaP_affine_new}
\end{align}

Equation~\eqref{eqn:DeltaP_affine_new} can be written compactly over the horizon as
\begin{align}
    \bm{\Delta P}_{\mathrm{agg}} = \tilde{\mathrm{A}}\,\bm{\tilde{\alpha}},
    \label{eqn:DeltaP_linear_map_new}
\end{align}
where bold notation is used to denote vectors, with
\begin{align}
    \bm{\tilde{\alpha}}
    &:=
    \begin{bmatrix}
        \tilde{\alpha}[0] & \tilde{\alpha}[1] & \cdots & \tilde{\alpha}[K-1]
    \end{bmatrix}^{\top},\\
    \bm{\Delta P}_{\mathrm{agg}}
    &:=
    \begin{bmatrix}
        \Delta P_{\mathrm{agg}}[1] & \Delta P_{\mathrm{agg}}[2] & \cdots & \Delta P_{\mathrm{agg}}[K]
    \end{bmatrix}^{\top},
\end{align}
and the entries of $\tilde{\mathrm{A}}$ are given by
\begin{align}
    a_{ij}:=
    \begin{cases}
        h_{i,j-1}, & j \le i,\\
        0, & j > i.
    \end{cases}
\end{align}

The matrix $\tilde{\mathrm{A}}$ depends only on the model dynamics, the initial conditions, and the forecasted ambient-temperature trajectory, and can therefore be computed offline. 

\subsection{Reduced Optimization Formulation}

Using the linear mapping in~\eqref{eqn:DeltaP_linear_map_new}, the original intractable optimization problem can be replaced by a reduced linear program.

For a positive setpoint change, the boundary of the reach-and-hold set is obtained from
\begin{subequations}\label{eqn:Opt_tractable_alpha_reduced_pos}
\begin{align}
    \max_{\bm{\tilde{\alpha}},\,t} \qquad & t \\
    \text{s.t.} \qquad 
    & \bm{\Delta P}_{\mathrm{agg}} = \tilde{\mathrm{A}}\,\bm{\tilde{\alpha}} \label{eqn:reduced_map_constraint_pos}\\
    & t \le \Delta P_{\mathrm{agg}}[k], \qquad \forall k\in\mathcal{K}_{\mathrm{hold}} \\
    & \bm{\tilde{\alpha}} \succeq \mathbf{0}_K \label{eqn:alpha_nonnegative_pos} \\
    & \mathbf{1}_K^\top \bm{\tilde{\alpha}} \le \beta \label{eqn:alpha_less_than_sum}.
\end{align}
\end{subequations}
The corresponding formulation for a negative setpoint change is analogous and is omitted for brevity; it is obtained by replacing the maximization with a minimization and reversing the hold inequality.

The reduced formulation in~\eqref{eqn:Opt_tractable_alpha_reduced_pos} contains only $K+1$ scalar decision variables, one per time step, rather than the original high-dimensional control input $u[k]\in\mathbb{R}^{2N}$, significantly reducing the size of the optimization problem. 

After solving for $\tilde{\alpha}[k]$, the corresponding broadcast probability $\alpha[k]$ can be recovered from
\begin{align}
    \alpha[k]
    =
    \frac{\tilde{\alpha}[k]}{s[k]}
    =
    \frac{\tilde{\alpha}[k]}{\beta-\sum_{l=0}^{k-1}\tilde{\alpha}[l]},
    \label{eqn:alpha_recovery_new}
\end{align}
    whenever $s[k]>0$. If $s[k]=0$, then no devices remain in the nominal-setpoint group, so determining $\alpha[k]$ is unnecessary.

\begin{prop}
The reach-and-hold set characterized by \eqref{eqn:Opt_tractable_alpha_reduced_pos} is an inner approximation of the true reach-and-hold set.
\end{prop}

\begin{proof}
Let $\mathcal{R}_{\mathrm{in}}$ denote the reach-and-hold set characterized by the reduced formulation. From~\eqref{eqn:alpha_control_law_new}, if $\alpha[k]\in[0,1]$ for all $k\in\mathcal{K}$, then the resulting control input $u[k]$ is admissible for the original problem.

Now, feasibility of the reduced formulation implies $\tilde{\alpha}[k]\ge 0$ for all $k$ and $\mathbf{1}_K^\top\bm{\tilde{\alpha}}\le \beta$. Hence,
\begin{align}
    0 \le \tilde{\alpha}[k] \le \beta-\sum_{l=0}^{k-1}\tilde{\alpha}[l] = s[k],
\end{align}
which, together with~\eqref{eqn:alpha_recovery_new}, implies $0\le \alpha[k]\le 1$. Thus, all feasible solutions to \eqref{eqn:Opt_tractable_alpha_reduced_pos} are also feasible solutions to \eqref{eqn:Opt_intractable_u[k]}. Hence, $\mathcal{R}_\text{in}\subseteq \mathcal{R}$.
\end{proof}

\begin{remark}[Heterogeneous considerations]\label{remark:heterogeneous considereations}
The reduced formulation~\eqref{eqn:Opt_tractable_alpha_reduced_pos} extends naturally to heterogeneous TCL populations with multiple parameter clusters and/or thermostat setpoints. If the fleet is partitioned into groups $g=1,\ldots,G$, each with its own representative parameter vector and homogeneous setpoint/deadband values, then each group has its own matrix $\tilde{\mathrm{A}}^{(g)}$ and input sequence $\bm{\tilde{\alpha}}^{(g)}$. The group dynamics are decoupled, and coupling arises only through the aggregate power deviation,
\[
\Delta \bm{P}_{\mathrm{agg}}
=
\sum_{g=1}^{G}\tilde{\mathrm{A}}^{(g)}\bm{\tilde{\alpha}}^{(g)}.
\]
Thus, compared to the single-cluster case, the formulation simply introduces one input sequence per group while preserving the same LP structure. In implementation, the scalar broadcast signal becomes a broadcast vector with one component per group.
\end{remark}

\subsection{Reach and Hold Characterization}
\label{sec: RnH_Characterization}

We now use the reduced LP to characterize the reach-and-hold set and examine key aspects of the framework.
We consider a heterogeneous population of 10,000 HVAC units governed by the second-order ETP model~\eqref{eq:ETP_2D}, with parameters sampled from the distributions in Table~\ref{tab:TCL Parameters}. The Markov model uses a discretization of $r=40$ bins$/^\circ$C and the ambient temperature profile in Fig.~\ref{fig:T_amb}. Both the Markov model and the agent-based simulation are initialized at the start of the control horizon $\mathcal{K}$ with the steady-state population temperature distribution corresponding to $T_{\mathrm{amb}}[0]$.

    \begin{table}[t]
        \centering
        \caption{ETP Parameters}
        \begin{tabular}{ccc}
        \toprule
        Parameter & Distribution & Units \\
        \midrule
        $C_\text{a}$ & $\mathcal{U}(0.51,0.63)$ & kWh$/^\circ$C \\
        $C_\text{m}$ & $\mathcal{U}(2.03,2.48)$ & kWh$/^\circ$C \\
        $U_\text{a}$ & $\mathcal{U}(0.25,0.30)$ & $^\circ$C/kW \\
        $H_\text{m}$ & $\mathcal{U}(4.43,5.41)$ & $^\circ$C/kW \\
        $P_\text{rated}$ & $\mathcal{U}(4.81,5.88)$ & kW \\
        \bottomrule
        \end{tabular}
        \label{tab:TCL Parameters}
    \end{table}
    
\begin{figure}[!t]
    \centering
    \includegraphics[width=0.95\linewidth]{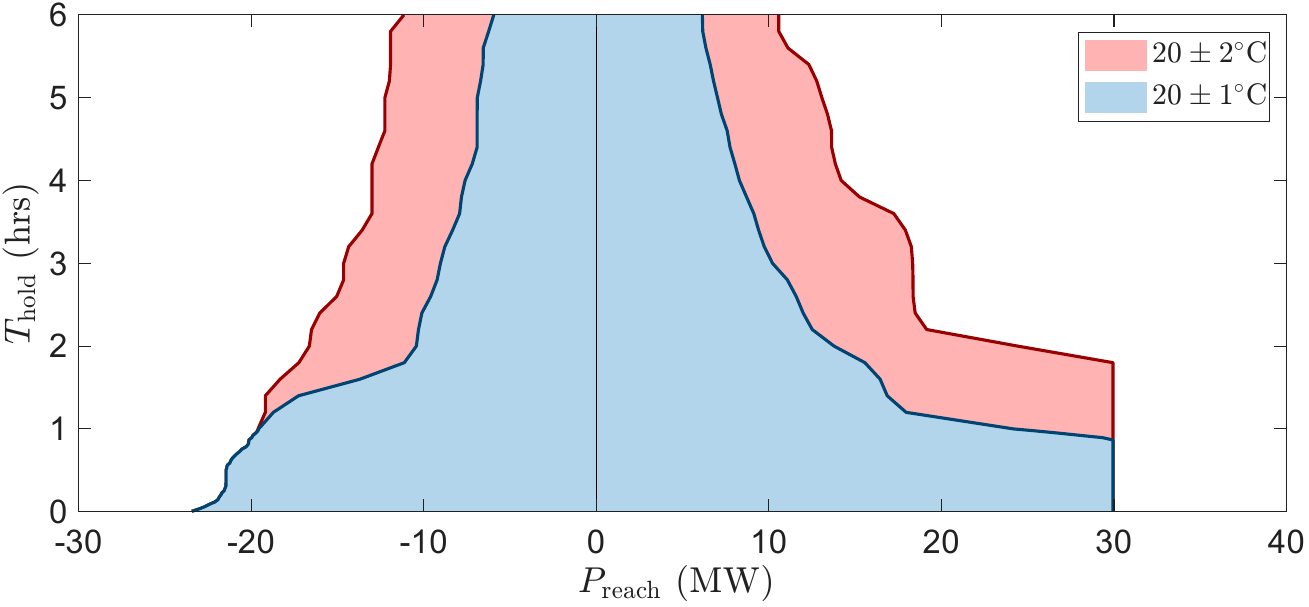}
   \caption{Reach-and-hold sets for setpoint changes of $20\pm1^\circ$C and $20\pm2^\circ$C.}
    \label{fig:RnH_base}
\end{figure}

We begin by characterizing the reach-and-hold set using~\eqref{eqn:Opt_tractable_alpha_reduced_pos} for $20\pm1^\circ$C and $20\pm2^\circ$C setpoint changes. Specifically, the boundary of each set is obtained by sweeping the hold duration $T_{\mathrm{hold}}\in[0,6]$ hours and, for each value, solving the reduced formulation to determine the corresponding maximum achievable value of $P_{\mathrm{reach}}$.

Fig.~\ref{fig:RnH_base} shows the resulting reach-and-hold sets. As expected, the $\pm2^\circ$C setpoint change yields a larger reach-and-hold set than the $\pm1^\circ$C case. However, both cases act on the same population, so the maximum achievable $P_{\mathrm{reach}}$ is the same and corresponds to switching essentially the entire available population to the DR setpoint. Consequently, the two sets coincide at small values of $T_{\mathrm{hold}}$. Similar to the battery example in Fig.~\ref{fig:Battery_Rnh}, the TCL reach-and-hold sets exhibit the basic tradeoff between $P_{\mathrm{reach}}$ and $T_{\mathrm{hold}}$. Unlike the smooth battery case, however, the TCL sets are less regular due to the fleet's non-zero nominal power consumption and the time-varying ambient conditions that shape the underlying transient dynamics.

\subsection{Peak Limiting Constraint to Limit Transient Peaks}

\begin{figure}[!t]
    \centering
    \includegraphics[width=0.95\linewidth]{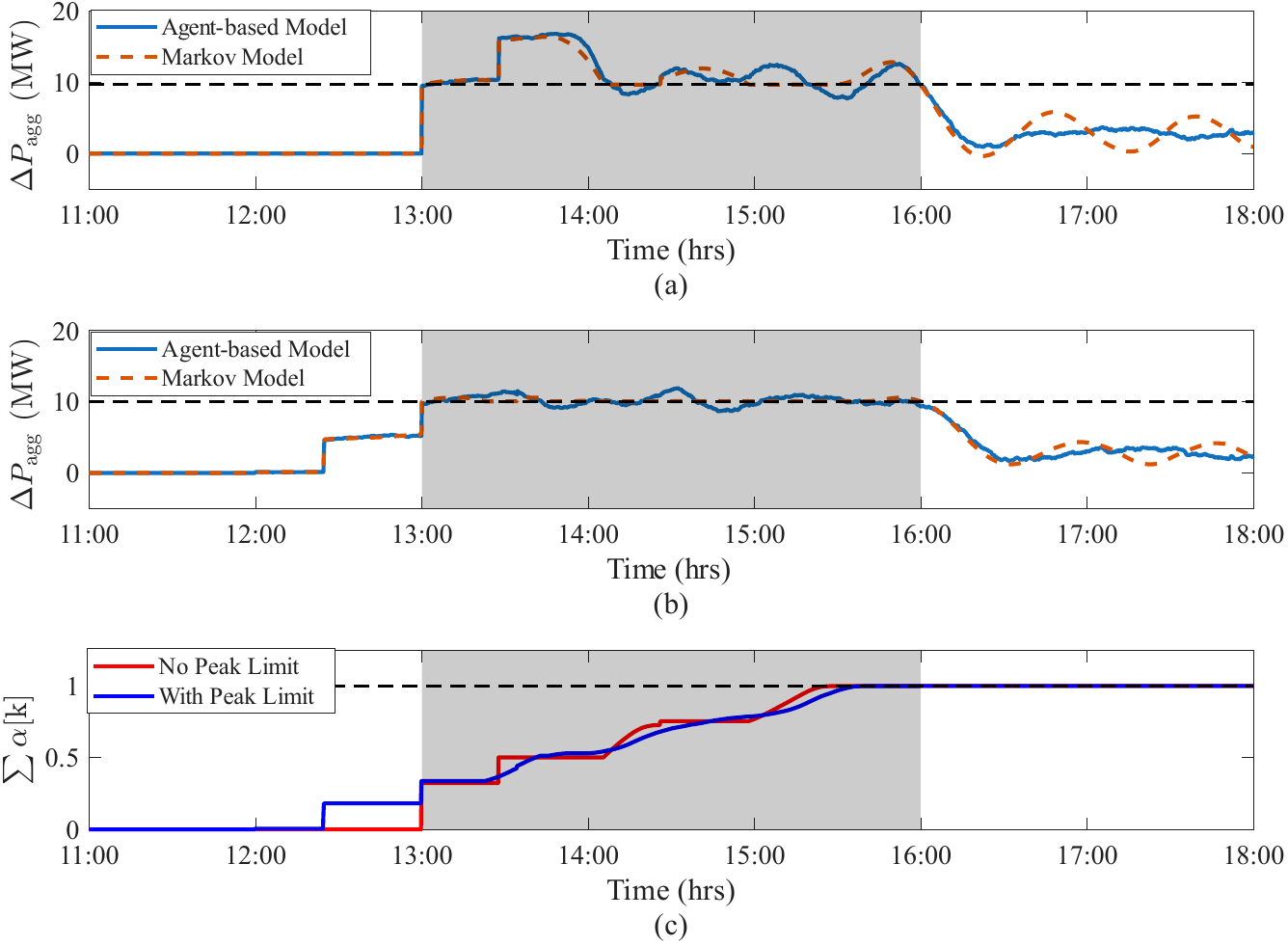}
 \caption{Open-loop implementations for the reach-and-hold point $(P_\text{reach},T_\text{hold})=(10,3)$ with and without peak limiting. (a) Aggregate power trajectory $\Delta P_{\mathrm{agg}}[k]$ for the unconstrained case. (b) Aggregate power trajectory $\Delta P_{\mathrm{agg}}[k]$ for the peak-limited case with $\gamma=0.05$. (c) Cumulative sum of $\tilde{\alpha}[k]$ for both cases}

    \label{fig:RnH_peak_limited_traj}
\end{figure}

The reduced LP in~\eqref{eqn:Opt_tractable_alpha_reduced_pos} maximizes the sustained reach value over the hold window. However, the resulting aggregate response may exhibit transient peaks above the maintained level. This is illustrated in Fig.~\ref{fig:RnH_peak_limited_traj}(a), which shows the open-loop implementation for the reach-and-hold point $(P_\text{reach},T_\text{hold})=(10,3)$. While the desired $P_\text{reach}$ is achieved in the Markov model, the trajectory exhibits noticeable transient peaks above $P_\text{reach}$.

To limit such behavior, we augment the reduced formulation in~\eqref{eqn:Opt_tractable_alpha_reduced_pos} with the additional constraint
\begin{align}
    \Delta P_{\mathrm{agg}}[k] \le (1+\gamma)t, 
    \qquad \forall k\in\mathcal{K}_{\mathrm{hold}},
    \label{eqn:peak_limiting_constraint}
\end{align}
where $\gamma>0$ is a tuning parameter. At optimality, $t=P_{\mathrm{reach}}$, so~\eqref{eqn:peak_limiting_constraint} requires $\Delta P_{\mathrm{agg}}[k]$ to remain within a factor of $(1+\gamma)$ of $P_{\mathrm{reach}}$ over the hold window. Smaller values of $\gamma$ therefore enforce a flatter response. The negative setpoint-change case is modified analogously.

Fig.~\ref{fig:RnH_peak_limited_traj}(b) shows the corresponding trajectory for the same reach-and-hold point when the peak limiting constraint is enforced with $\gamma=0.05$. Compared to the case without the peak limiting constraint, the resulting trajectory is noticeably flatter while still achieving the desired reach-and-hold point in the Markov model. To further illustrate how this flatter response is achieved, Fig.~\ref{fig:RnH_peak_limited_traj}(c) plots the cumulative sum of the input $\tilde{\alpha}[k]$ for both cases, which represents the total fraction of devices that have switched to the DR setpoint up to time $k$. Without the peak limiting constraint, the cumulative input exhibits larger jumps, indicating that larger fractions of the population are switched at a few time instants, which produces sharper transient peaks in $\Delta P_{\mathrm{agg}}[k]$. In contrast, under the peak limiting constraint, the cumulative input evolves more gradually and begins earlier, showing that the open-loop schedule staggers the switching of devices over time to avoid large transient peaks.

\begin{figure}[!t]
    \centering
    \includegraphics[width=0.95\linewidth]{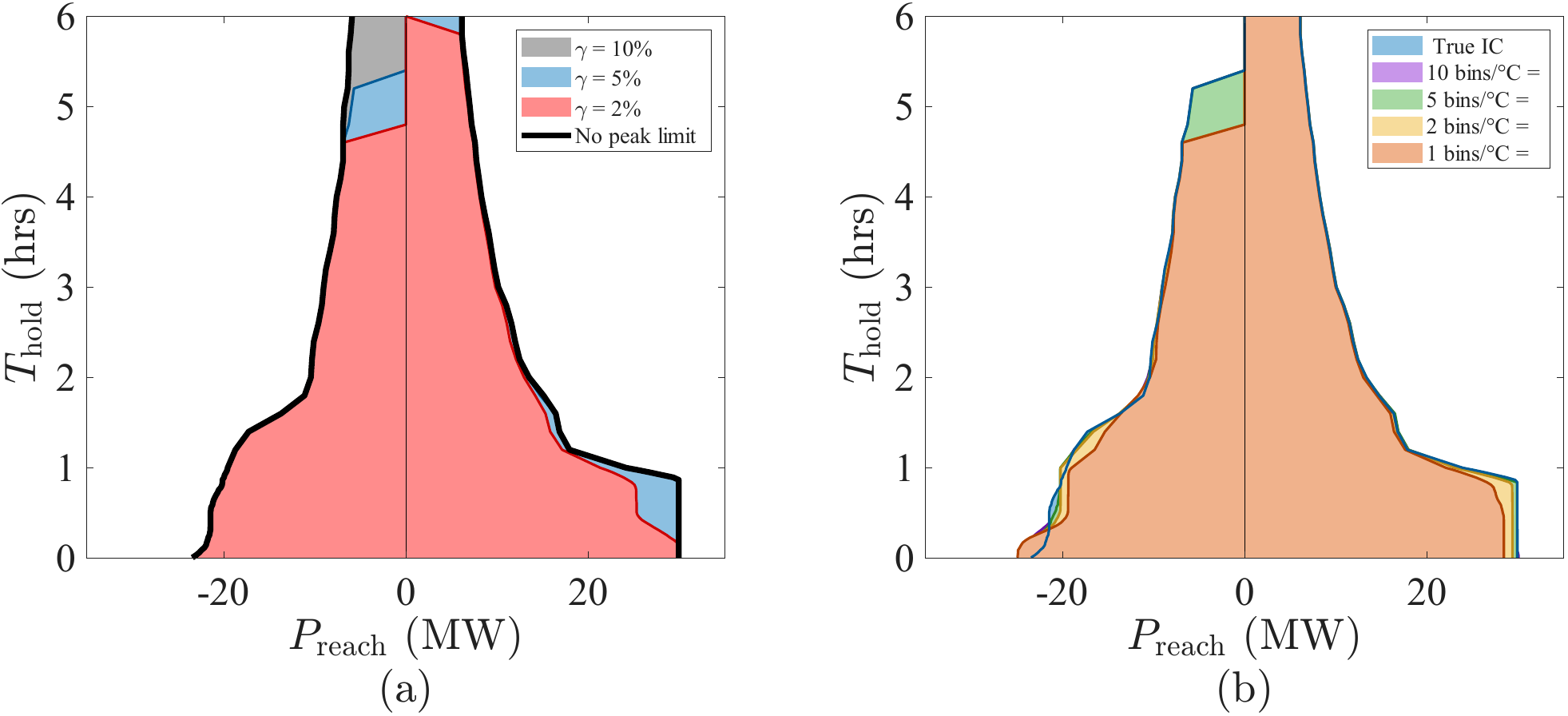}

    \caption{Reach-and-hold sets for a $\pm1^\circ$C setpoint change: (a) with different peak limiting parameters $\gamma$, and (b) with different coarse initial-condition estimates.}
    \label{fig:RnH_smooth_set}
\end{figure}

To examine the effect of the peak limiting constraint on the characterized reach and hold set,  we solve the peak-limited formulation for $T_{\mathrm{hold}}\in[0,6]$ hours and for different values of $\gamma$. For comparison, we also solve~\eqref{eqn:Opt_tractable_alpha_reduced_pos} without the peak limiting constraint. Fig.~\ref{fig:RnH_smooth_set}(a) shows the resulting reach-and-hold sets for a $\pm1^\circ$C setpoint change. 

The characterized reach-and-hold set shrinks as $\gamma$ decreases. For $\gamma \ge 0.1$, the reduction in set size is relatively small, whereas for $\gamma \le 0.05$ the set contracts more noticeably. Furthermore, for sufficiently small $\gamma$, portions of the set collapse toward $P_{\mathrm{reach}}=0$ at longer hold durations. This occurs because time-varying ambient conditions cause the transition matrices $A_{\mathrm{dr}}[k]$ to vary over the hold window, producing a non-uniform aggregate response to the setpoint change. As a result, sustaining a nonzero $P_{\mathrm{reach}}$ over long durations inherently produces transient peaks in $\Delta P_{\mathrm{agg}}[k]$ above the sustained level. When $\gamma$ is small enough, the peak limiting constraint prohibits these peaks, rendering all nonzero $P_{\mathrm{reach}}$ values infeasible. The effect is more pronounced for negative setpoint changes, since the nominal power trajectory starts to decrease after about 5~hours, and additional setpoint switching beyond that point increases power consumption, thereby further increasing the peak of $\Delta P_{\mathrm{agg}}$. Thus, the choice of $\gamma$ depends on the application and the grid and DER operator's desired aggregate power trajectories.

In Fig.~\ref{fig:RnH_peak_limited_traj}(a)--(b), the Markov-model trajectories achieve the desired $P_\text{reach}$ as per Definition~\ref{def:reach-hold}, whereas the agent-based simulation shows deviations due to modeling mismatch. This realized discrepancy is studied next.

\section{Validation of the Reach-and-Hold Set}\label{sec: Simulation Results}
The reach-and-hold characterization developed in the previous sections, together with the corresponding open-loop schedules, is obtained using the aggregate Markov model. A natural next question is how accurate are these characterizations when the open-loop schedules are implemented on a more detailed agent-based model of the TCL fleet. This section therefore evaluates the realized performance of the proposed schedules and then examines the robustness of the resulting reach-and-hold characterization to uncertainty in initial conditions and TCL parameters.

\subsection{Open-loop Implementation}
The open-loop schedules $\bm{\Tilde{\alpha}}$ obtained for $\gamma = 0.05$, a $+1 ^\circ$C setpoint change, and $T_\text{hold} = \{1,3,5\}$~hours are applied to the agent-based simulation. The resulting $\bm{\Delta P}_\text{agg}$ trajectory is assessed in terms of the corresponding $P_\text{reach}$ value. Additionally, to further assess performance over different operational timescales, we compute the realized $P_\text{reach}$ value using averaging windows of varying durations to capture the typical practical timescales related to measurement and verification of demand response programs~\cite{CAISO_FLEXmeter_2022}. Consider an averaging window of duration $\delta$ with corresponding window length $L(\delta) := \delta/\Delta t \in \mathbb{N}$ that partitions $\mathcal{K}_\text{hold}$ into $N_\delta$ contiguous, non-overlapping index sets $\{\mathcal{I}_j\}_{j=1}^{N_\delta}$\footnote{Here, $\mathcal{I}_j:=\{k_0+(j-1)L(\delta),\dots,k_0+\min(jL(\delta)-1,|\mathcal{K}_\text{hold}|-1)\}$ and $N_\delta:=\lceil |\mathcal{K}_\text{hold}|/L(\delta) \rceil$.}.

For each interval $\mathcal{I}_j$, we compute the $\delta$-averaged aggregate power and define the corresponding $\delta$-averaged realized reach value as
\begin{align}
    P_\text{reach}^\delta
    :=
    \min_{j=1,\dots,N_\delta}
    \left\{
        \frac{1}{|\mathcal{I}_j|}
        \sum_{k\in\mathcal{I}_j}\Delta P_\text{agg}[k]
    \right\}.
\end{align}
We then quantify the mismatch between the predicted and realized $P_\text{reach}$ values
\begin{align}
    \varepsilon(\delta):=
    \left|
    \frac{P_{\text{reach}}^\delta-P_{\text{reach}}^{\ast}}
         {P_{\text{reach}}^{\ast}}
    \right|\times 100\%,
\end{align}

where $P_\text{reach}^{\ast}$ denotes the predicted reach value from the characterized reach-and-hold set. In the results below, we report $\varepsilon(\delta)$ for $\delta\in\{$10-sec, 5-min, 15-min, 60-min, $T_\text{hold}\}$. For $\delta=T_\text{hold}$, the averaging window spans the full DR event. These averaged metrics also provide insight into how short-term deviations dissipate. The resulting percentage errors are listed in Table~\ref{tab:reach-and-hold Results}.

\begin{table}[!t]
    \centering
    \caption{Open-loop reach-and-hold results for $+1 ^\circ$C setpoint change.}
    \begin{tabular}{cccccc}
  \toprule
      $T_\text{hold}$ &  $\varepsilon(10~\text{sec})$  & $\varepsilon(5~\text{min})$  & $\varepsilon(15~\text{min})$  &$\varepsilon(60~\text{min})$  & $\varepsilon(T_\text{hold})$  \\
    (hrs) & &  & & &  \\
    \midrule
    1 & 3.82\% & 2.92\% & 1.96\% & 0.12\% & 0.12\%  \\
    3 & 10.1\% & 8.27\% & 4.35\% & 0.98\% & 0.19\%  \\
    5 & 11.6\% & 9.21\% & 6.16\% & 1.18\% & 0.93\%  \\
    \bottomrule
    \end{tabular}
    \label{tab:reach-and-hold Results}
\end{table}
From Table~\ref{tab:reach-and-hold Results}, the realized $P_\text{reach}$ shows reasonably good agreement with the values predicted by the Markov model. The averaged results show that the error decreases noticeably as the averaging window increases, indicating that much of the mismatch is driven by short-term deviations in the realized trajectory. Thus, although the instantaneous response can deviate from the characterized $P_\text{reach}$ value, the realized aggregate response aligns more closely with the predicted behavior when viewed over longer timescales.

\subsection{Robustness Analysis}

We next investigate the effects of relaxing \textbf{Assumptions~1} and~\textbf{2} on the reach-and-hold set and open-loop performance.

\subsubsection{Relaxing Assumption~1 (Initial Conditions)}

 In Section~\ref{sec: RnH}, we assumed that the initial distributions $x_{\mathrm{n},0}$ and $x_{\mathrm{dr},0}$ were known exactly. However, this assumption is impractical. Participating thermostats can provide indoor air-temperature measurements, but typically only at coarse resolution, on the order of $0.1-0.5^\circ $C~\cite{resideo_th303}. Additionally, the building-mass temperature is generally not directly measurable. The goal of this subsection is therefore to quantify the sensitivity of the characterized set and the resulting open-loop schedules to errors in the initial population state.

\begin{figure}
    \centering
    \includegraphics[width=\linewidth]{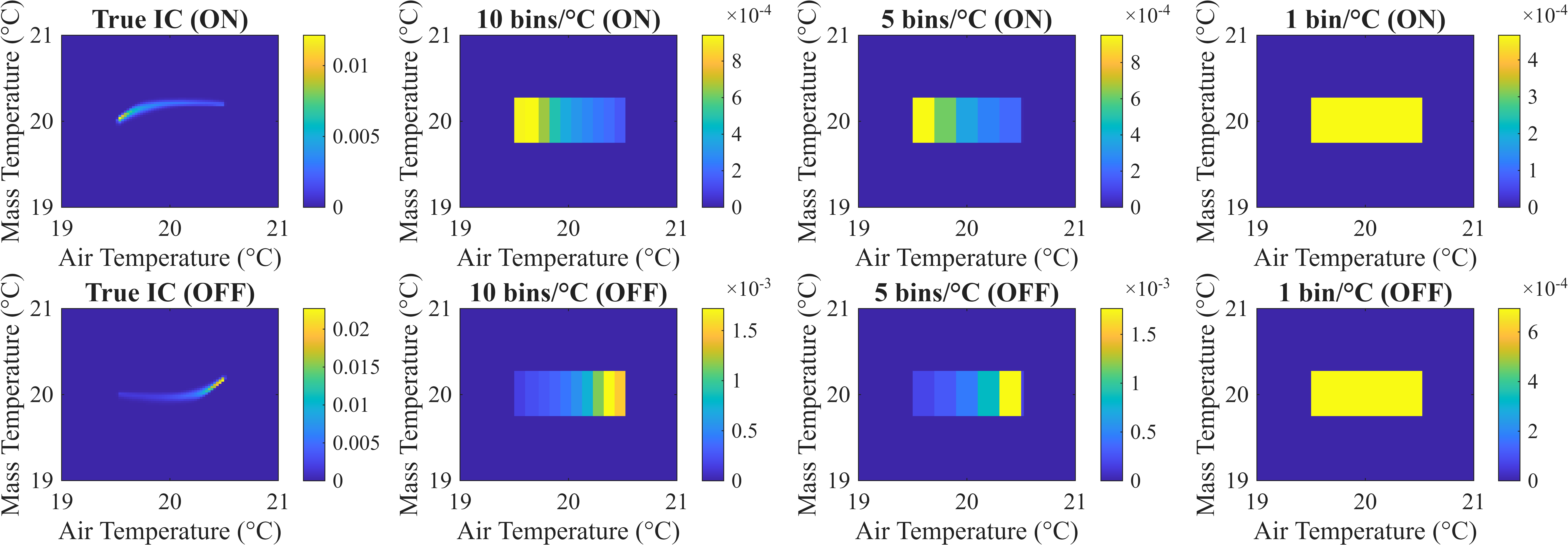}
    \caption{True and coarse 2-D initial-condition estimates for the ON and OFF populations. The four columns correspond to the true distribution, a 10 bins/°C estimate, 5 bins/°C estimate, and a 1 bin/°C estimate.}
    \label{fig:IC_heatmap}
\end{figure}

To assess the effect of this limited observability, we perform a simulation-based sensitivity study in which the true initial condition is replaced by coarse initial-condition estimates. Specifically, we consider a sequence of increasingly coarse estimates of the initial air-temperature distribution, with resolutions ranging from 10 bins/$^\circ $C down to 1 bin/$^\circ $C. For each coarse air-temperature estimate, the corresponding mass-temperature distribution is not assumed to be directly observed. Instead, we construct an approximate initial condition by distributing the population uniformly in the mass-temperature dimension over the interval
$[T_\text{set}-\frac{\Delta T}{4},T_\text{set}+\frac{\Delta T}{4}]$. This interval reflects the slower evolution of the mass temperature relative to the indoor air temperature. The chosen interval is not intended to exactly reconstruct the unmeasured mass-temperature distribution, but rather to test the sensitivity of the proposed reach-and-hold characterization to a plausible coarse approximation of this unobserved mass temperature. The resulting coarse estimates are then mapped onto the fixed 40 bins/$^\circ$C discretization used in the Markov model. Fig.~\ref{fig:IC_heatmap} illustrates the true initial condition together with the coarse estimates.

Using each estimated initial condition, we characterize the corresponding reach-and-hold sets with the peak limiting constraint $\gamma=0.05$, and compare its boundary with the corresponding sets obtained using the true initial condition. The resulting sets are shown in Fig.~\ref{fig:RnH_smooth_set}(b). For each case the maximum absolute deviation in the boundary value of $P_\text{reach}$ across all considered hold durations was computed. The largest deviations occur for the coarsest initial-condition estimates, with errors of approximately $6.3$~MW for resolutions of 1 and 2 bins/$^\circ$C. For resolutions of 5 bins/$^\circ$C or higher, the maximum deviation drops below $1.7$~MW, indicating that moderate air-temperature resolution is sufficient to substantially reduce the sensitivity of the characterized set to initialization errors.

To evaluate the effect of these initialization errors on realized performance, we compute the corresponding reach-and-hold set and associated open-loop schedule $\bm{\tilde{\alpha}}_i$ for each estimated coarse initial condition $i$. The resulting schedules are then applied to the agent-based simulation initialized with the true initial condition, and the corresponding aggregate power trajectories $\Delta P_{\mathrm{agg},i}[k]$ are recorded for each case $i$. Using the same metric as before, we then compute the realized reach value $P_{\text{reach},i}^{\delta}$ and corresponding error $\varepsilon_i(\delta)$ for each initial-condition $i$. Table~\ref{tab:reach-and-hold Results IC} reports the resulting values of $\varepsilon_i(\delta)$ for the different averaging windows and initial-condition resolutions. This process is applied for $T_\text{hold}=3$ hrs and assessed for $\delta\in\{$10-sec, 5-min, 15-min, 60-min, $T_\text{hold}\}$. For $\delta=T_\text{hold}$, the averaging window spans the full DR event.

\begin{table}[!t]
    \centering
    \caption{Performance under initial-condition uncertainty for $T_\text{hold} = 3$hrs.}
    \begin{tabular}{cccccc}
    \toprule
    Resolution &  $\varepsilon(10~\text{sec})$  & $\varepsilon(5~\text{min})$  & $\varepsilon(15~\text{min})$  &$\varepsilon(60~\text{min})$  & $\varepsilon(T_\text{hold})$  \\
    (bins/$^\circ$C) & &  & &  &  \\

    \midrule
        1  & 16.7\%  &14.8\% &9.81\%    & 1.43\%  & 1.28\% \\
    5  & 13.9\%  &10.7\% &8.27\%    & 1.37\% & 0.52\% \\
    10 & 12.1\%  &10.4\% & 6.22\%   & 1.17\% & 0.32\% \\
    \bottomrule
    \end{tabular}
    \label{tab:reach-and-hold Results IC}
\end{table}

The results indicate that uncertainty in the initial condition mainly affects short-timescale, instantaneous power deviations, as reflected by the larger errors in $P_{\mathrm{reach}}$ at very small averaging intervals. However, these errors decay rapidly with increasing averaging window. In particular, when performance is evaluated over 15-minute windows, the realized response remains within 10\% of the predicted response for all three initial-condition estimates. Additionally, when the study was repeated for $T_\text{hold}=\{1,3,5\}$ hours, the same conclusion was observed for 15-minute windows.

\subsubsection{Relaxing Assumption~2 (ETP Parameter Values)}
In characterizing the reach-and-hold set, we also assumed that the ETP parameters of the TCLs were known, which might not be the case in practice. To investigate the effect of parameter uncertainty, open-loop $\bm{\Tilde{\alpha}}$ schedules obtained using the nominal parameters listed in Table~\ref{tab:TCL Parameters} were implemented on an agent-based simulation where the parameters of the TCLs were perturbed from their nominal values. ETP model parameters were estimated within 5\% error in~\cite{31_Parameter_Identifiability} and therefore we perturb the parameters by $\pm5\%$. We apply the perturbations on $U_\text{a},C_\text{a},H_\text{m}$ and $ C_\text{m}$ for a total of 81 ($3^4$) combinations. The values of $P_\text{rated}$ and $\eta$ are assumed to be known and can be obtained from the nameplate of the TCLs. 

\begin{figure}[!t]
    \centering
    \includegraphics[width=0.9\linewidth]{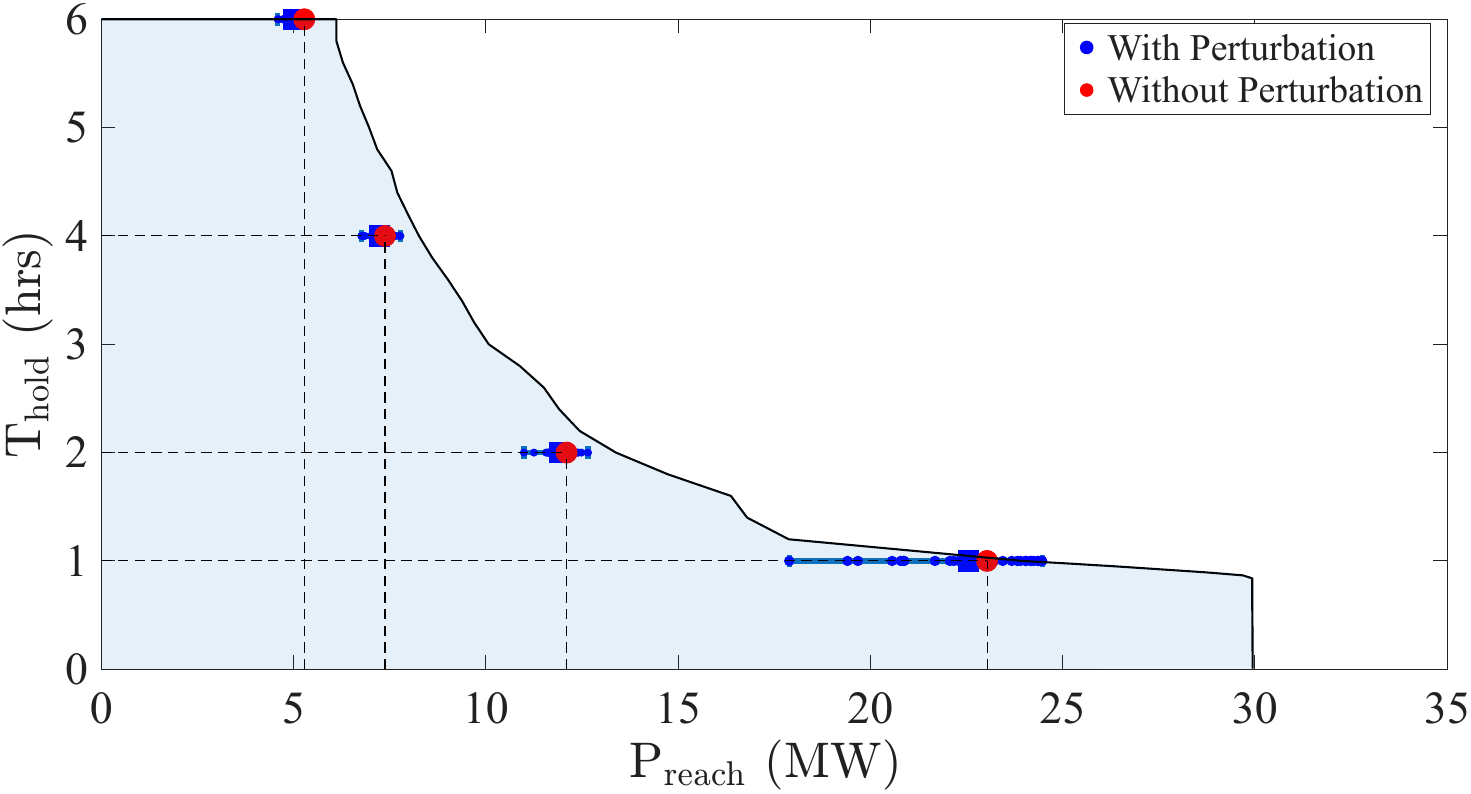}
    \caption{Realized $P_\text{reach}$ values from open-loop implementation of $\bm{\Tilde{\alpha}}$ schedule obtained with and without parameter perturbation. The mean of all the parameter perturbation cases is illustrated using the blue square. }
    \label{fig:Realized_Preach_dA}
\end{figure}

The open-loop $\bm{\Tilde{\alpha}}$ schedule obtained for $T_\text{hold} = \{1,2,4,6\}$~hours for a $+1 ^\circ$C setpoint change using the nominal parameters was implemented on all 81 perturbed agent-based simulations. Fig.~\ref{fig:Realized_Preach_dA} shows the realized $(P_\text{reach},T_\text{hold})$ pairs with and without parameter variation. 

The performance degradation caused by parameter changes is quantified in Table~\ref{tab:reach-and-hold Results Params} for different averaging windows and $T_\text{hold}$ values, where $\varepsilon_\text{max}(\delta)$ denotes the maximum error across all parameter-perturbation cases.
\begin{table}[!t]
    \centering
    \caption{Performance under ETP parameter uncertainty.}
\begin{tabular}{cccccc}
    \toprule
    $T_\text{hold}$ & $\varepsilon_\text{max}(\delta)$  & $\varepsilon_\text{max}(\delta)$  & $\varepsilon_\text{max}(\delta)$  &$\varepsilon_\text{max}(\delta)$  & $\varepsilon_\text{max}(\delta)$ \\
    (hrs) & (10-sec)  &  (5-min) & (15-min)&  (60-min)& ($T_\text{hold}$)  \\
    \midrule
    1 & 22.3\% & 17.5\% & 11.7\% & 4.83\% & 4.83\% \\
    2 & 17.8\% & 15.9\% &   13.7\% & 6.40\% & 5.71\% \\
    4 & 18.1\% & 15.2\% &  13.9\% &6.28\% & 4.88\% \\
    6 & 20.9\% & 18.3\% &  15.8\% & 6.72\% & 4.66\% \\
    \bottomrule
    \end{tabular}
    \label{tab:reach-and-hold Results Params}
\end{table}
The results show that the parameter perturbation has a significant effect on the instantaneous power values with $P_\text{reach}$ errors reaching~22.3\%. However, when the average power trajectory is considered the errors reduce significantly, indicating that the instantaneous power errors are not sustained over longer durations.
Overall, these validation results show that, for performance windows of 15 minutes or longer, the proposed reach-and-hold framework generally predicts realized open-loop performance within 15\% across the uncertainty cases and hold durations considered, and within about 6\% when averaged over the entire event. This provides DER aggregators and grid operators with a practical tool for characterizing the power-duration capability of aggregated TCL fleets under setpoint control.

\section{Conclusion}\label{sec:Conc}
This paper introduced a computationally tractable framework to characterize the flexibility of aggregated TCLs under thermostat setpoint control in terms of a reach-and-hold set, which provides an \emph{ex-ante} power-duration characterization of fleet capability. A Markov-chain aggregate model was used to capture TCL population dynamics, and an equivalent reduced formulation was developed to obtain a tractable LP-based inner approximation of the reach-and-hold set together with associated open-loop broadcast schedules.
 The open-loop control schedules obtained from the optimization formulation were shown to accurately steer the aggregate power consumption to achieve the required $P_\text{reach}$ and $T_\text{hold}$ when simulated on aggregate models and agent-based simulations. This work provides grid and DER operators with a practical tool for optimizing DR scheduling. Future research will focus on robustness against parameter and ambient temperature uncertainties and on quantifying the conservativeness of the proposed inner approximation. Additionally, we are interested in incorporating user (dis)comfort to better capture how end-user effects impact the available flexibility for DR.

\renewcommand*{\bibfont}{\footnotesize}
\printbibliography
\end{document}